\documentclass[useAMS,usenatbib]{mn2e}
\usepackage[dvips]{graphicx}
\usepackage[dvips]{color}
\usepackage{longtable}
\usepackage{float}

\def\lesssim{\mathrel{\hbox{\rlap{\hbox{\lower4pt\hbox{$\sim$}}}\hbox{$<$}}}}
\def\gtrsim{\mathrel{\hbox{\rlap{\hbox{\lower4pt\hbox{$\sim$}}}\hbox{$>$}}}}
\newcommand{\mincir}{\raise-2.truept\hbox{\rlap{\hbox{$\sim$}}\raise5.truept
\hbox{$<$}\ }}
\newcommand{\magcir}{\raise-2.truept\hbox{\rlap{\hbox{$\sim$}}\raise5.truept
\hbox{$>$}\ }}
\newcommand{\be}{\begin{equation}}
\newcommand{\ee}{\end{equation}}
\newcommand{\ba}{\begin{eqnarray}}
\newcommand{\ea}{\end{eqnarray}}

\newcommand{\HII}{{H}{II}}

\newcommand{\muh}{\mu_{\mathrm{H}}}
\newcommand{\nuh}{\nu_{\mathrm{H}}}

\title[ShaSS: Ram-pressure stripping vs. tidal interaction]{Shapley
  Supercluster Survey: Ram-Pressure Stripping vs. Tidal Interactions
  in the Shapley Supercluster \thanks{Based on data collected with i)
    WiFeS at the 2.3m telescope of the Australian National University
    at Siding Spring (Australia) and ii) OmegaCAM at the ESO INAF - VLT
    Survey Telescope and VIRCAM at VISTA, both at the European
    Southern Observatory, Chile (ESO Programmes 088.A-4008,
    089.A-0095, 090.A-0094, 091.A-0050, 093.A-0465).}}

\author[Merluzzi et al.]{P. Merluzzi$^{1}$, G. Busarello$^{1}$,
  M. A. Dopita$^{2,3}$, C.~P. Haines$^{4,5}$, D. Steinhauser$^{6}$,
  \newauthor H. Bourdin$^{7,8}$, P. Mazzotta$^{7}$
  \\ merluzzi@na.astro.it \\ $^1$ INAF-Osservatorio Astronomico di
  Capodimonte, Via Moiariello 16 I-80131 Napoli, Italy \\ $^2$
  Research School of Astronomy and Astrophysics, Australian National
  University, Cotter Rd., Weston ACT 2611, Australia \\ $^3$ Astronomy
  Department, Faculty of Science, King Abdulaziz University, PO Box
  80203, Jeddah, Saudi Arabia \\ $^4$ Departamento de Astronom\'{i}a,
  Universidad de Chile, Casilla 36-D, Correo Central, Santiago,
  Chile\\ $^5$ INAF-Osservatorio Astronomico di Brera, Via Brera 28
  I-20121 Milano, Italy\\ $^6$ Institute of Astro- and Particle
  Physics, University of Innsbruck, Technikerstr. 25, 6020 Innsbruck,
  Austria \\ $^7$ Dipartimento di Fisica, Universit\`a di Roma Tor
  Vergata, Via della Ricerca Scientifica 1, I-00133 Roma, Italy\\ $^8$
  Harvard-Smithsonian Center for Astrophysics, 60 Garden St,
  Cambridge, MA 02138, United States}

\begin{document}

\date{Accepted . Received }

\pagerange{\pageref{firstpage}--\pageref{lastpage}} \pubyear{}

\maketitle

\label{firstpage}

\begin{abstract}
We present two new examples of galaxies undergoing transformation in
the Shapley supercluster core. These low-mass
($\mathcal{M}_{\star}\sim 0.4-1\times10^{10}$M$_\odot$) galaxies are
members of the two clusters SC\,1329-313 ($z\sim0.045$) and
SC\,1327-312 ($z\sim0.049$). Integral-field spectroscopy complemented
by imaging in $ugriK$ bands and in H$\alpha$ narrow-band are used to
disentangle the effects of tidal interaction (TI) and ram-pressure
stripping (RPS). In both galaxies, SOS\,61086 and SOS\,90630, we
observe one-sided extraplanar ionised gas extending respectively
$\sim30$\,kpc and $\sim 41$\,kpc in projection from their disks. The
galaxies' gaseous disks are truncated and the kinematics of the
stellar and gas components are decoupled, supporting the RPS
scenario. The emission of the ionised gas extends in the direction of
a possible companion for both galaxies suggesting a TI. The overall
gas velocity field of SOS\,61086 is reproduced by {\it ad hoc}
N-body/hydrodynamical simulations of RPS acting almost face-on and
starting $\sim 250$\,Myr ago, consistent with the age of the young
stellar populations. A link between the observed gas stripping and the
cluster-cluster interaction experienced by SC\,1329-313 and A\,3562 is
suggested. Simulations of ram pressure acting almost edge-on are
  able to fully reproduce the gas velocity field of SOS\,90630, but
  cannot at the same time reproduce the extended tail of outflowing
  gas. This suggests that an additional disturbance from a TI is
  required. This study adds a piece of evidence that RPS may take
place in different environments with different impacts and witnesses
the possible effect of cluster-cluster merger on RPS.
\end{abstract}


\section{Introduction}
\label{intro}

\begin{figure*}
\includegraphics[width=180mm]{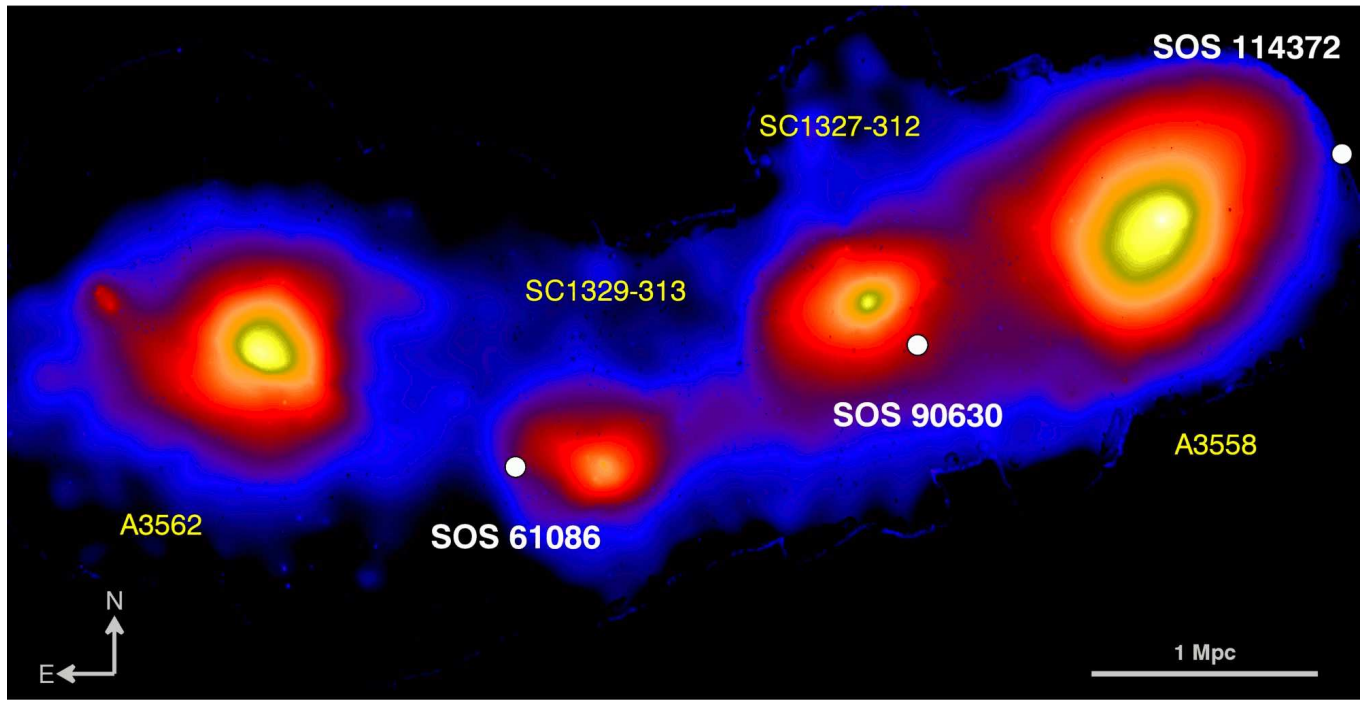}
\caption{The positions of SOS\,61086 and SOS\,90630 (white dots) are
  indicated on the X-ray emission map including four out of five
  clusters of the SSCC. The position of SOS\,114372 (see text) is also
  marked. The X-ray surface brightness is derived from a wavelet
  analysis of XMM-Newton images extracted in the $0.5-2.5$\,keV energy
  band, that have been corrected for spatially variable effective
  exposure and background components. Scale and orientation are shown
  in the bottom.}
\label{galsX}
\end{figure*}

The evolution of galaxies depends both on their intrinsic properties,
such as their mass, and external `accidents' in which the
galaxies are randomly involved during their life. These external
events include tidal interactions, galaxy mergers, ram-pressure and
viscous stripping, evaporation and `starvation'
\citep{TT72,MKL96,BV90,BH91,B01,GG72,N82,CS77,LTC80}. Together, these
events serve to transform galaxies by disturbing their kinematics,
depleting their reservoirs of gas, and so quenching star
formation. The time-scales and efficiencies of such mechanisms are
different and depend on both the properties of the galaxies and on
their environment \citep[for a review see][]{BG06}. In particular,
within the crowded and dense environments of galaxy clusters and
groups, the probability that hydrodynamical events transform galaxies
increases significantly. This may serve to explain the origin of the
different galaxy populations observed in clusters and field in the
local Universe \citep[e.g.][]{D80,L02,BNB09}.

In this work we investigate the possibility of disentangling two
mechanisms which have been invoked as main drivers in transforming
spiral galaxies into S0s and dEs: ram-pressure stripping
\citep[RPS,][]{GG72,AMB99} and galaxy-galaxy tidal interactions
\citep[TIs,][]{TT72,MKL96}.

Galaxies orbiting into a cluster feel the ram pressure exerted by the
hot and dense intracluster medium (ICM) which can effectively remove
the cooler interstellar medium (ISM) in the galaxy starting from
outside and thus quenching star formation in the ram-pressure stripped
regions. The time-scales for RPS is about one cluster crossing time
($\sim10^9$\,yr). As emphasised by \citet{GG72} this cluster-specific
mechanism efficiently depletes the gas of massive spiral galaxies only
within the cluster cores where the dense ICM is expected to be
located. However, the effects of ram pressure depend on both galaxy
and ICM properties and may easily extend to poorer environments for
low-mass galaxies \citep{MBD03}, and not only for them. Using
hydrodynamical cosmological simulations, \citet{BMB13} investigated
the increase of gas content and star formation in cluster galaxies
with the clustercentric distance. They found that this observed
large-scale trend, approaching the values of the field galaxy sample
only at $\sim 5r_{200}$, can be explained by a combination of i)
pre-processing’ of galaxies within infalling groups; ii)
‘overshooting’ for those galaxies that are not falling in for the
first time; and iii) ram-pressure stripping.  Simulations of ongoing
RPS show that such a mechanism not only severely truncates the gas
disk of $L^\star$ galaxies in high-density environments, but also in
low-density environments, where moderate ram pressures are foreseen,
their gas disk may be disturbed and bent \citep{RH05}.  Observations
of special events of gas stripping confirmed that RPS is acting more
efficiently than previously postulated by \citet{GG72}, playing a role
also outside the cluster cores \citep[e.g.][]{CGK07,ACCESSV}.

Ram pressure may also compress and shock the ISM, temporarily enhancing
the star formation in the inner disk as well as in the stripped gas
\citep{BV90,FN99,TB12,B14}. These effects are observed in a few cases
\citep{ACCESSV,KGJ14,ESE14}. Since it is highly effective in quenching
the star formation, RPS is also considered to be one of the most
important processes, although not the only one, to convert spirals and
irregulars into S0 and spheroidal galaxies \citep[see][]{KB12}.

\begin{table*}
  \centering
\caption{Properties of SOS\,61086 and SOS\,90630 and their
    neighbouring galaxies.}
{\small
\begin{tabular}{lcccc}
\hline
{\bf Property}  & {\bf SOS\,61086} & {\bf SOS\,61087} & {\bf SOS\,90630} & {\bf  SOS\,90090} \\
\hline
{\bf Coordinates} &&&& \\
ShaSS$^1$ & 13\,31\,59.80\,-31\,49\,22.2 & 13\,31\,59.85\,-31\,49\,04.0 &
13\,29\,28.53\,-31\,39\,25.6 & 13\,29\,23.43\,-31\,39\,51.7 \\
&&&& \\
{\bf Magnitudes/fluxes} &&&& \\
$u^{a,1}$ & 17.81${\pm}$0.03 & 20.06${\pm}$0.03 & 17.41${\pm}$0.03 & 17.58${\pm}$0.03 \\
$g^{a,1}$ & 16.62${\pm}$0.03 & 18.36${\pm}$0.03 & 16.39${\pm}$0.03 & 15.60${\pm}$0.03 \\
$r^{a,1}$ & 16.30${\pm}$0.03 & 17.44${\pm}$0.03 & 16.06${\pm}$0.03 & 14.70${\pm}$0.03 \\
$i^{a,1}$ & 16.23${\pm}$0.03 & 17.23${\pm}$0.03 & 15.96${\pm}$0.03 & 14.43${\pm}$0.03 \\
24$\mu$m$^2$ & 4007${\pm}$222\,$\mu$Jy & &11013${\pm}$573\,$\mu$Jy & 3136${\pm}$180\,$\mu$Jy \\
1.4GHz$^3$ & 0.89\,mJy & & 2.30\,mJy & \\
&&&& \\
{\bf Masses} &&&& \\
stellar mass$^{4}$ & $3.61\times10^{9}$M$_\odot$ & $4.85\times10^{9}$M$_\odot$ & $1.0\times10^{10}$M$_\odot$ & $8.72\times10^{10}$M$_\odot$ \\
total halo mass$^{4}$ & $1.9\times10^{11}$M$_\odot$ & $2.20\times10^{11}$M$_\odot$ & $2.5\times10^{11}$M$_\odot$ & $5.0\times10^{11}$M$_\odot$ \\
&&&& \\
{\bf Distances} &&&& \\
redshift$^{1}$ & 0.04261$\pm$0.00023 & 0.04367$\pm$0.00031 & 0.04817$\pm$0.00042 & 0.04929$\pm$0.00022 \\
heliocentric velocity$^{1}$ & 12502$\pm$69\,km\,s$^{-1}$ & 12806$\pm$93\,km\,s$^{-1}$ & 14093\,km\,s$^{-1}$ & 14413$\pm$66\,km\,s$^{-1}$ \\
projected distance$^{4}$ & 282\,kpc && 226\,kpc & \\
to the parent cluster centre &&&& \\
&&&& \\
{\bf Star formation rates} &&&& \\
UV+IR global SFR$^{2}$ & 1.4\,M$_\odot$yr$^{-1}$ && 2.5\,M$_\odot$yr$^{-1}$ & 0.5\,M$_\odot$yr$^{-1}$ \\
H$\alpha$ global SFR$^{4}$  & $1.76\pm 0.56$\,M$_\odot$yr$^{-1}$ && $3.49\pm 1.07$\,M$_\odot$yr$^{-1}$& \\
&&&& \\
\hline
\end{tabular}}
\begin{tabular}{l}
$a$: Magnitudes in the AB photometric system.\\
Sources: $^1$ \citet{ShaSSI}; $^2$ \citet{ACCESSII}; $^3$ \citet{M05}; $^4$ this work. \\
\end{tabular}
\label{gals}
\end{table*}

A counter-argument to this hypothesis is that S0s differ from normal
spirals by their higher bulge luminosities rather than simply having
fainter disks \citep{CZ04} and this is not explained with the RPS or
starvation mechanisms. Other mechanisms such as TIs \citep[merging,
  fly-bys;][]{TT72} and harassment \citep[which is also a TI in a
  wider sense;][]{MKL96} are better capable of channelling material
onto a central bulge, sufficient to produce the higher central mass
densities seen in cluster spirals and ultimately the stellar phase
densities found in S0s \citep{MMT07}. In general, TIs gravitationally
perturb the gas and stellar components, trigger central starbursts and
strip stars and gas from the disk of the involved spiral galaxies
producing tails and bridges
\citep[see][]{TT72,LT78,KK84,KKH85,BH91,BH92,BH96,WGB06}.

In cluster environments, due to the high relative velocities of the
colliding galaxies, TI time-scales are shorter ($\sim 10^8$yr) than in
the field, and a single passage may only marginally affect the
dynamics of their stellar populations. The geometry of the encounter
and the relative galaxy properties are important parameters in
determining the size of the dynamical perturbation. Multiple
encounters (harassment) -- which are of course more common in clusters
-- are probably necessary to substantially perturb the stellar
content.

The presence of both gas and stars outside the galaxy disk may arise
from either RPS or through TIs. Although each mechanism will produce
distinctive distributions in the extraplanar gas and stars, it may be
not straightforward to distinguish which process is acting on a
particular galaxy because projection effects may mask the geometries
of the outflowing matter \citep[e.g][]{BDA04,RB06}. The internal
kinematics can help to distinguish between these mechanisms; while the
gas component suffers the ram pressure which is able to modify its
velocity field, the stellar kinematics will be affected only in case
of TIs.

Most, if not all, these observables are transient phenomena. After
$\sim 1$\,Gyr the `memory' of either the RPS or TI is washed out
\citep[see][]{KKS06,OKL08}. The permanent results are i) the
star-formation quenching, achieved with both mechanisms, although on
different time-scales; ii) the truncation of the gaseous disks by RPS;
iii) the structural modification by TIs. Each galaxy can be involved
in different `accidents' across cosmic time, making it extremely
difficult to understand which mechanism really prevails in
transforming the star-forming disk galaxies accreted from the field
into the S0 and dE populations which now dominate cluster cores.

To address the challenge of understanding the physics of galaxy
transformations, we need integral-field spectroscopy (IFS). IFS allows
us to resolve the different spatial components in the galaxies, and so
measure the dynamical disturbance of the stellar component, discover
disturbed gas velocity fields, determine the local enhancement and
spatial trend of the star formation, and to use chemical abundance
analysis to identify the source of the extraplanar gas.

With this in mind, we have undertaken IFS observations with WiFeS
\citep{Dopita07,Dopita10} of a small but carefully selected galaxy
sample in the Shapley supercluster core (SSCC), drawn from the Shapley
Supercluster Survey \citep[ShaSS][]{ShaSSI} which aims to investigate
the role of the mass assembly on galaxy evolution. It covers
23\,deg$^2$ centred on the SSCC with ESO-VST $ugri$ and ESO-VISTA
$K$-band imaging. The Shapley supercluster is located at a redshift
$z\sim 0.05$, and represents the most dynamically-active and dense
structure in the local Universe. Thus, the probability to observe
evidence of environmental effects on galaxy evolution is dramatically
enhanced. The targets for the IFS have been selected from the
spectroscopic catalogue of ShaSS which is 80 per cent complete down to
$i = 17.6$ (m$^\star$+3). All these galaxies are supercluster members,
resolved in the optical images and display i) disturbed morphologies,
such as asymmetry and tails; ii) hints of extraplanar emission; iii)
evidence of star-forming knots. About 80 galaxies satisfy at least two
of these properties. After this {\it visual} selection the galaxies
are targeted with a 45-minute exposures of WiFeS which allow to
ascertain which of them actually present extraplanar emission and then
become the high-priority targets in our investigation. At present 17
supercluster galaxies have been observed. They belong to different
environments, from dense cluster cores to the regions where
cluster-cluster interactions are taking place, and out to the much
less populated areas.

In our first study \citep{ACCESSV} we identified a bright
(L$>$L$^\star$) barred spiral galaxy (SOS\,114372) $\sim$1\,Mpc from
the centre of the rich cluster A\,3558 in the SSCC, which is being
affected by RPS. IFS observations revealed ongoing gas stripping in
the form of one-sided extraplanar ionised gas along the full extent of
the disk, simultaneously with a starburst triggered by gas compression
along the leading edge of the galaxy. The galaxy is subjected to
weak-moderate ram pressure, as defined by \citet{RH05}. This adds a
piece of evidence to the fact that RPS is acting more efficiently on
the galaxy ISM than previously foreseen and also outside of the
cluster cores, as also observed in the Virgo cluster by \citet{CGK09}.
This is possibly the principal transformation process quenching star
formation (SF) in spirals, although certainly it is helped by other
processes affecting the structure of the galaxies.

In this work we present the results for the galaxies SOS\,61086 and
SOS\,90630. These two galaxies are 2.9 and 2.3 $K$-band magnitudes
fainter with respect to SOS\,114372 allowing us to extend the
investigation of the environmental effects to a lower stellar mass
range, but also to investigate different environments, since both are
members of low-mass clusters ($\mathcal{M}_{cl}\sim 10^{14}$M$_\odot$) involved
in interactions.

In Sect.~\ref{targets} the main properties of the targets are
presented. Observations and data reduction are described in
Sects.~\ref{IFS} and \ref{mmtf}. The data analysis is summarized in
Sect.\ref{DA}. The results are given in Sects.~\ref{RES61086} and
\ref{RES90630} for the two galaxies. In Sect.~\ref{epg}, we discuss
the possible origins of the observed extraplanar gas and in
Sec.~\ref{sim} we compare the observed gas velocity field with those
derived from hydrodynamical simulations and discuss possible drivers
for the observed transformations. Our conclusions are summarized in
Sec~\ref{summary}.

Throughout the paper, and in common with the other papers of this
series, we adopt a cosmology with $\Omega_M$=0.3, $\Omega_\Lambda$=
0.7, and H$_0$=70\,km\,s$^{-1}$Mpc$^{-1}$. According to this cosmology
1\,arcsesc corresponds to 0.880\,kpc at $z$=0.0447 (SC\,1329-313) and
0.966\,kpc at $z$=0.0493 (SC\,1327-312) . Velocities and velocity
dispersions are given accounting for the relativistic correction.

\begin{figure}
\includegraphics[width=80mm]{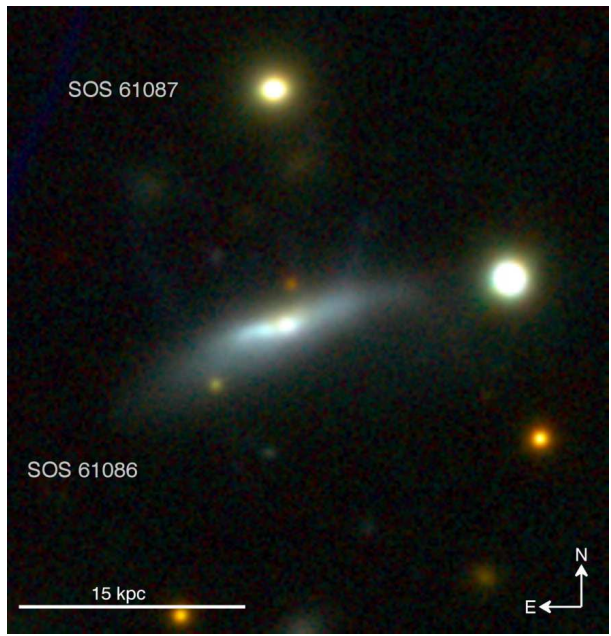}
\caption{Composite $gri$ image of the field including SOS\,61086
  in the centre and SOS\,61087 located North.}
\label{SOS61086VST}
\end{figure}

\begin{figure}
\includegraphics[width=80mm]{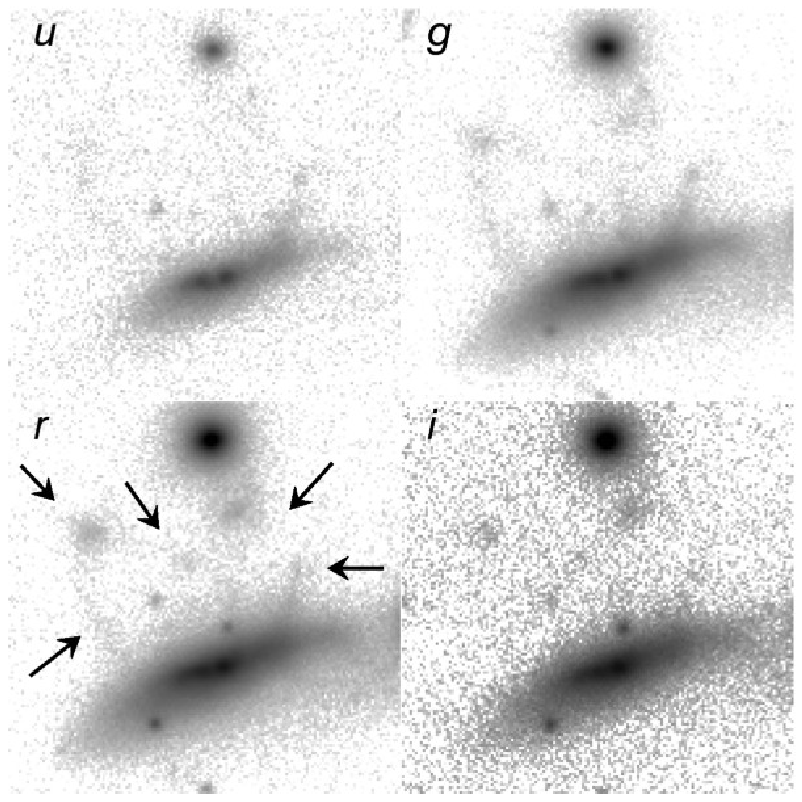}
\caption{Detail of the field of Fig.~\ref{SOS61086VST} shown in $ugri$
  bands, with hints of (low-brightness) extraplanar material
  highlighted by the arrows on the $r$-band image.}
\label{SOS61086VST_4}
\end{figure}

\begin{figure}
\includegraphics[width=80mm]{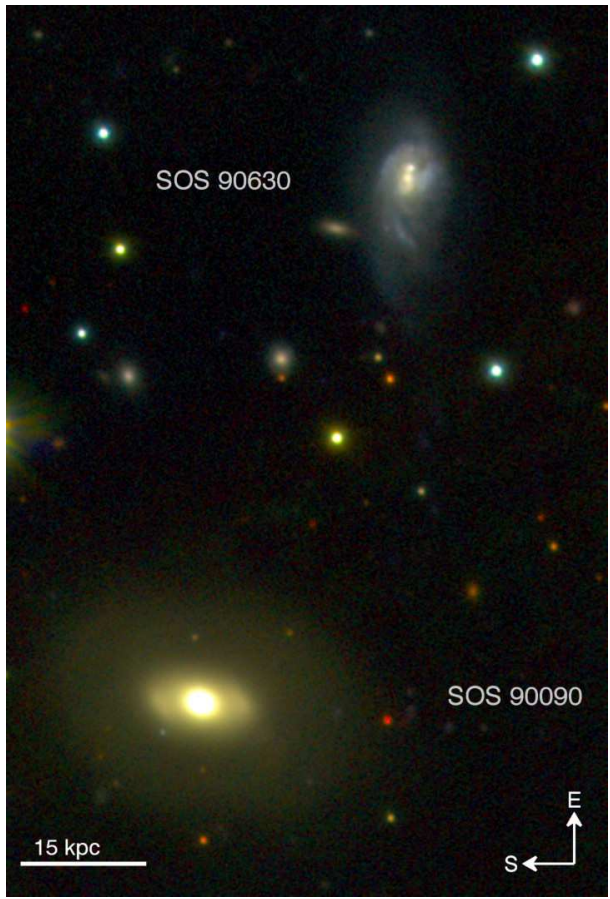}
\caption{Composite $gri$ image of the field including SOS\,90630 (top)
  and SOS\,90090 (bottom).}
\label{SOS90630VST}
\end{figure}

\begin{figure}
\includegraphics[width=80mm]{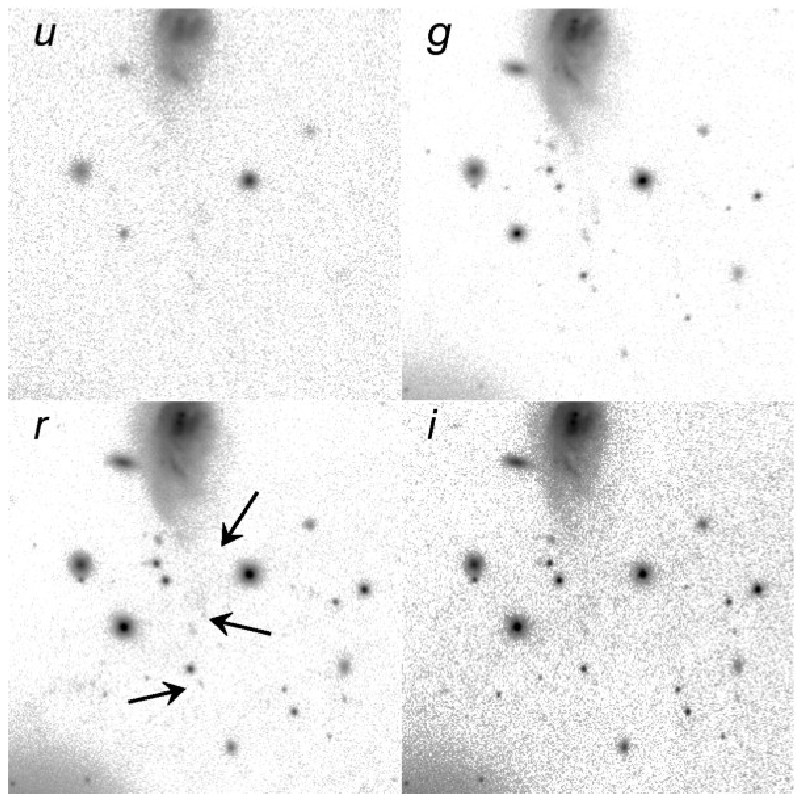}
\caption{Detail of the field of Fig.~\ref{SOS90630VST} shown in $ugri$
  bands, with hints of (low-brightness) extraplanar material
  highlighted by the arrows on the $r$-band image.}
\label{SOS90630VST_4}
\end{figure}

\section{The galaxies}
\label{targets}

The galaxies SOS\,61086 and SOS\,90630, named following the Shapley
Optical Survey (SOS) identification\footnote{In the Shapley
  supercluster survey \citep[ShaSS,][]{ShaSSI} they correspond to the
  galaxies ShaSS403015754 and ShaSS408053848, respectively
  \citep[][]{ShaSSII}.} \citep{SOSI,SOSII}, are spiral galaxies
respectively 2.7 and 2.1 magnitudes fainter than m$^\star$ in $K$ band
\citep[$K^\star$=11.7 at the supercluster redshift,][]{ACCESSI} and
with stellar masses $\mathcal{M}_ {\star}\sim
0.4\times10^{10}$M$_\odot$ and $\mathcal{M}_ {\star}\sim
1\times10^{10}$M$_\odot$. Both lie in poor cluster environments, as
indicated in Fig.~\ref{galsX} (white dots) relative to the SSCC as
traced by the X-ray surface brightness derived by XMM-Newton images in
the 0.5-2.5\,keV energy band (see Sect.~\ref{ICMpro}).

\subsection{SOS\,61086} 
\label{SOS61086des}

This galaxy has a redshift $z$=0.04261 ($V_{h}=12502$\,km\,s$^{-1}$)
and is located at 282\,kpc in projection from the X-ray centre of the
cluster SC\,1329-313 ($\mathcal{M}_{cl}\sim
3.7\times10^{14}$M$_\odot$, Ragone et al. \citeyear{RMP06}) in the
SSCC. SC\,1329-313 forms a well-defined, distinct structure both
spatially and in velocity space. The distribution of galaxies and the
X-ray emission are both elongated along the ENE direction towards
Abell\,3562 \citep[see also][]{FHB04,GVB05}. By analysing the caustic
diagram of the ShaSS spectroscopic sample (Haines et al., in
preparation), we determine its mean recession velocity of
13114\,km\,s$^{-1}$ ($z=0.04475$), with a velocity dispersion of just
$348{\pm}28$\, km\,s$^{-1}$, based on 55 member galaxies within
$r_{200}$ (0.86\,Mpc, see also Bardelli et al. \citeyear{BPR98}). Both
dominant elliptical galaxies have very similar recession velocities
(12807 and 12758\,km\,s$^{-1}$), which are 300-350\,km\,s$^{-1}$ lower
than the mean. This suggests that the line-of-sight (LOS) peculiar
velocity of SOS\,61086 with respect to the main group is about
-600\,km\,s$^{-1}$.

A $gri$ composite image of the galaxy, derived from the optical
imaging of ShaSS, is shown in Fig.~\ref{SOS61086VST}. The pixel scale
of the imaging is 0.214\,arcsec/pxl with a seeing less than
0.8\,arcsec in $gr$ bands and $\sim0.9$\,arcsec in $i$
band. Fig.~\ref{SOS61086VST} shows SOS\,61086 in the field centre and
another cluster member, SOS\,61087, at about 17\,kpc to the North. The
bright source to the right of SOS\,61086 is a star.
Fig.~\ref{SOS61086VST_4} shows part of the field of
Fig.~\ref{SOS61086VST} in the four VST bands $ugri$. It focuses on the
area where hints of extraplanar material are present (highlighted by
the arrows in the r-band image). These figures show hints of matter
beyond the stellar disk in the northerly direction with respect to
SOS\,61086. In particular, two faint filaments leading NW and NE from
the disk few kiloparsecs in projection from the galaxy centre. The
extraplanar emission seems to form a fan-like structure bordered by
the two filaments. Two faint clumps located NW and NE with respect to
the galaxy centre and out of the disk can be also distinguished. It is
not clear if these clumps are associated to the galaxy, the NW one
being redder with respect to the main galaxy body. It is interesting
to notice that although some of these extraplanar features may be very
faint, most of them are present at all wavebands.

SOS\,61086 is a disk galaxy seen almost edge-on showing some
distortion in the optical images. This is most probably due to an
irregular distribution of dust, since in $K$ band the galaxy looks
much more symmetrical (see Sect.~\ref{SF_1}).

SOS\,61086 could be affected by the presence of the close companion
galaxy SOS\,61087. The redshift of this neighbour is $z=0.04367$, implying a LOS
velocity of +304\,km\,s$^{-1}$ relative to SOS\,61086.  While this
companion galaxy appears much more compact than SOS\,61086, its
estimated stellar mass is 1.3$\times$ higher.

The main properties of SOS\,61086 and SOS\,61087 are listed in
Table~\ref{gals}. The Kron magnitudes in the table are corrected for
Galactic extinction following \citet{SF11}.

From the ultra-violet (UV) and mid-infrared (mid-IR) fluxes we
estimate a global star formation rate SFR=1.40$^{+0.26}_{-0.16}$\,M$_\odot$yr$^{-1}$
\citep[of which 47\% is obscured,][]{ACCESSII}. For the stellar mass we
adopted the calibration of the GAMA survey \citep{THB11}. For the
characterization of the dark matter halos we adopted the models by
\citet{DBD14}, which account for the dependence of the halo properties
on the central galaxy \citep[eq.~1 of][]{DBD14}.

\subsection{SOS\,90630} 

SOS\,90630, at redshift $z$=0.04817 ($V_{h}=14093$\,km\,s$^{-1}$), is
located at 226\,kpc in projection from the X-ray centre of the cluster
SC\,1327-312 ($\mathcal{M}_{cl}\sim 3\times10^{14}$M$_\odot$, Ragone
et al. \citeyear{RMP06}) in the SSCC (see Sect.~\ref{epg}). The centre
of SC1327-312 is well defined with the peak of X-ray emission
coinciding with a bright elliptical galaxy (6dF J1329477-313625,
$z=0.05017$, $V_{h}=14664$\,km\,s$^{-1}$). The central velocity of the
group is slightly lower at 14429\,km\,s$^{-1}$ ($z=0.04935$) and the
velocity dispersion is $535{\pm}17$\,km\,s$^{-1}$. This suggests a LOS
peculiar velocity of SOS\,90630 with respect to the main group of
about -300\,km\,s$^{-1}$.

The $gri$ composite VST image in Fig.~\ref{SOS90630VST} was derived
from images with seeing 0.6\,arcsec in $gr$ bands and 0.5\,arcsec in
$i$ band. The light distribution of SOS\,90630 is highly asymmetric
with a prominent western arm in the direction where the disk seems
more extended. Star-formation knots and dust obscured regions are
found in the centre. Hints of matter flowing out of SOS\,90630 towards
West are visible in Fig.~\ref{SOS90630VST_4} in all bands.

The main properties of SOS\,90630 and SOS\,90090 are listed in
Table~\ref{gals}. For SOS\,90630 we measured a global
SFR=2.50$^{+0.72}_{-0.44}$\,M$_\odot$yr$^{-1}$ \citep[of which 74\% is
obscured;][]{ACCESSII}. SOS\,90090 has a 5$\times$ lower SFR and
4.6$\times$ higher stellar mass (see Table~\ref{gals}). Using the same
parametrizations as for SOS\,61086, we estimate the stellar mass and
mass halo of SOS\,90630 listed in Table~\ref{gals}.

We notice that in the optical ($gri$) images SOS\,90630 presents two
peaks of brightness in the centre, but this is due to dust absorption
as demonstrated in Appendix~\ref{DN_app} and Sect.~\ref{RES90630}.

\begin{figure}
\includegraphics[width=80mm]{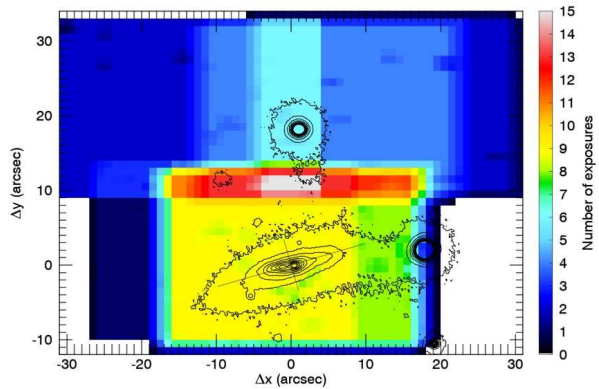}
\caption{{\bf SOS\,61086} WiFeS spatial coverage of the field
  including SOS\,61086. The number of exposures covering each part of
  the field are indicated in the coloured bar. In this figure and in
  the following maps the axes indicate the apparent distance from the
  $K$-band photometric centre in arcsecs, the curves are $r$-band
  isophotes and the two lines crossing the centre mark the apparent
  major and minor axes and extend to 3 disk scale radii. With around
  6-8 exposures we reach SNR=5 for a flux of
  0.5$\cdot$10$^{-17}$erg\,s$^{-1}$cm$^{-2}$\,arcsec$^{-2}$ for the
  H$\alpha$ line.}
\label{SOS61086COV}
\end{figure}

\begin{figure}
\includegraphics[width=80mm]{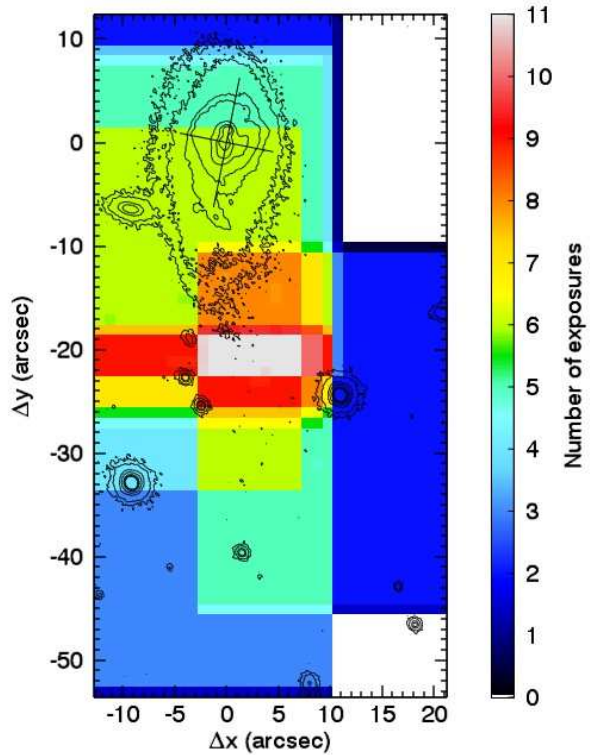}
\caption{{\bf SOS\,90630} Same as Fig.~\ref{SOS61086COV} for
  SOS\,90630.}
\label{SOS90630COV}
\end{figure}

\section{Integral-field spectroscopy}
\label{IFS}

The IFS data of SOS\,61086 and SOS\,90630 were obtained
during two observing runs in April 2011 and April 2012 using the
Wide-Field Spectrograph \citep[WiFeS][]{Dopita07,Dopita10} mounted at
the Nasmyth focus of the Australian National University 2.3m telescope
located at the Siding Spring Observatory, Australia.  WiFeS is an
image-slicing integral-field spectrograph that records optical spectra
over a contiguous 25$^{\prime\prime}$$\times$\,38$^{\prime\prime}$
field-of-view. This field is divided into twenty-five
1$^{\prime\prime}$-wide long-slits (`slices') of 3$8^{\prime\prime}$
length. WiFeS has two independent channels for the blue and the red
wavelength ranges. We used the B3000 and R3000 gratings, allowing
simultaneous observations of the spectral range from $\sim $3300\,\AA\
to $\sim $9300\,\AA\ with an average resolution of R=2900. For further
details on the WiFeS instrument see \citet{Dopita07,Dopita10}.

To cover the galaxies and their surroundings at sufficient depth, we
obtained 15 and 11 pointings (45\,min each) on SOS\,61086 and
SOS\,90630, respectively. The coverage maps of the two galaxies are
shown in Figs.~\ref{SOS61086COV} and \ref{SOS90630COV}.  For each
exposure on the target, we also acquired the spectrum of a nearby
empty sky region with 22.5\,min exposure to allow accurate sky
subtraction.

We obtained spectra of spectrophotometric standard stars for flux
calibration and B-type stars with nearly featureless spectra to
determine the telluric correction. Arc and bias frames were also taken
for each science exposure.  Internal lamp flat fields and sky flats
were taken twice during both runs.

The data were reduced using the WiFeS data reduction pipeline
\citep{Dopita10} and purposely written FORTRAN and
IDL\footnote{http://www.exelisvis.com/ProductsServices/IDL.aspx} codes
\citep[for details see][]{ACCESSV}. The WiFeS pipeline performs all
the steps from bias subtraction to the production of wavelength- and
flux-calibrated data-cubes for each of the blue and red channels. The
final spectral resolution achieved is $\sigma{\sim}40$\,km\,s$^{-1}$,
and is wavelength and position dependent. The accuracy of the
  wavelength calibration is 0.3\,\AA\footnote{This is the standard
    deviation of the difference between the observed wavelength and
    that predicted by the adopted dispersion relation, averaged over
    the 25x38 spectra of the datacubes.}. The data-cubes were sampled at
1$^{\prime\prime}$$\times$1$^{\prime\prime}$$\times$1\,\AA\ and cover
a useful wavelength range of 3800--8500\,\AA. The data-cubes were also
corrected for atmospheric differential refraction.  Sky subtraction
was carried out by means of sky spectra taken closest in time to the
galaxy spectra.

The co-addition of individual exposures/pointings was performed taking
into account the instrumental spatial distortion as detailed in
\citet{ACCESSV}. The reduced data-cubes were corrected for Galactic
extinction following \citet{SFD98} and using the extinction curve by
\citet{CCM89} with R$_V$=3.1. The sensitivity of our data is
0.5$\cdot$10$^{-17}$erg\,s$^{-1}$cm$^{-2}$\,arcsec$^{-2}$ at a
signal-to-noise ratio SNR=5 for the H$\alpha$ line, which is usually
achieved with 6-8 exposures of 45\,min each. Further details on data
reduction are given in \citet{ACCESSV}.

\begin{figure*}
\includegraphics[width=164mm]{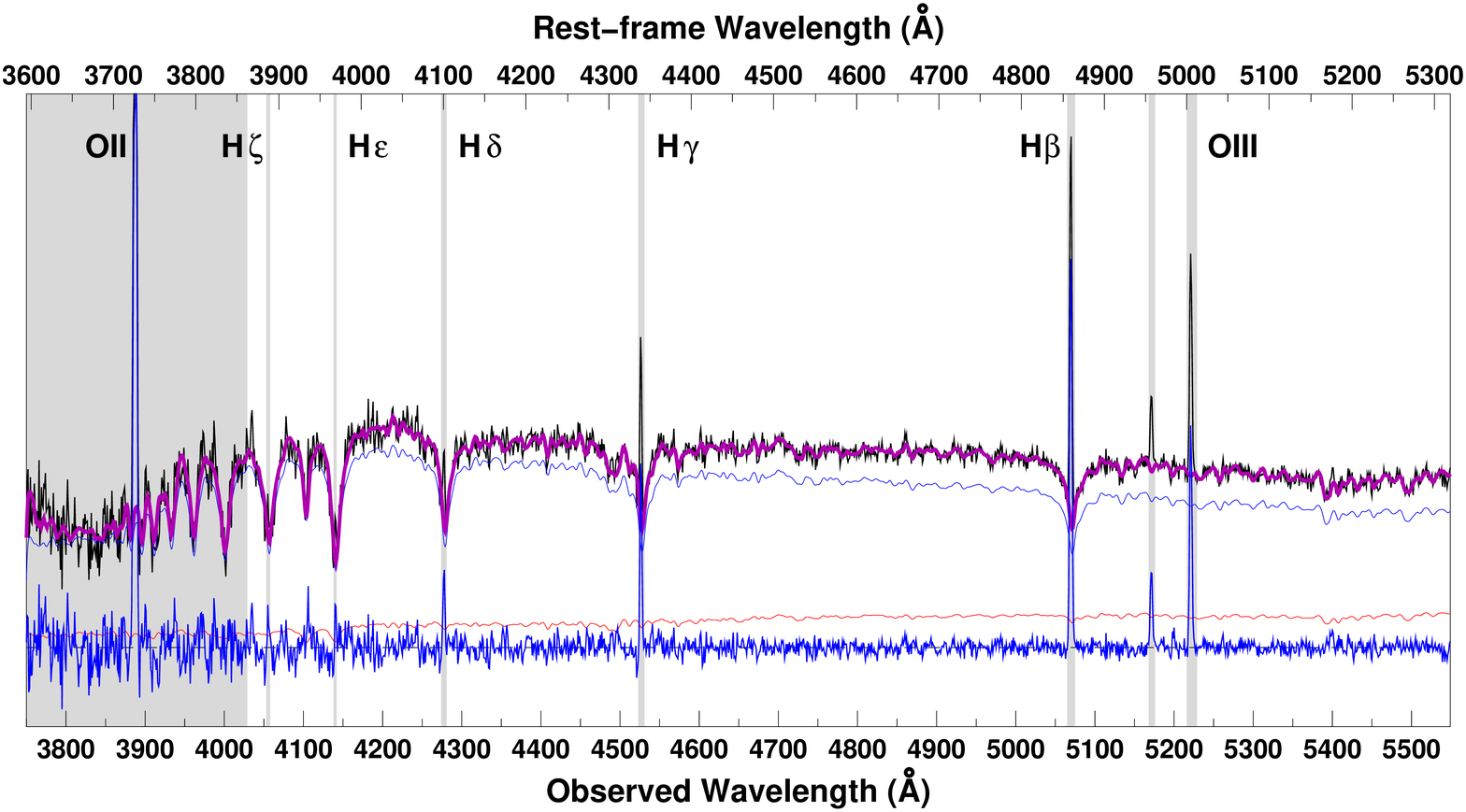}
\includegraphics[width=164mm]{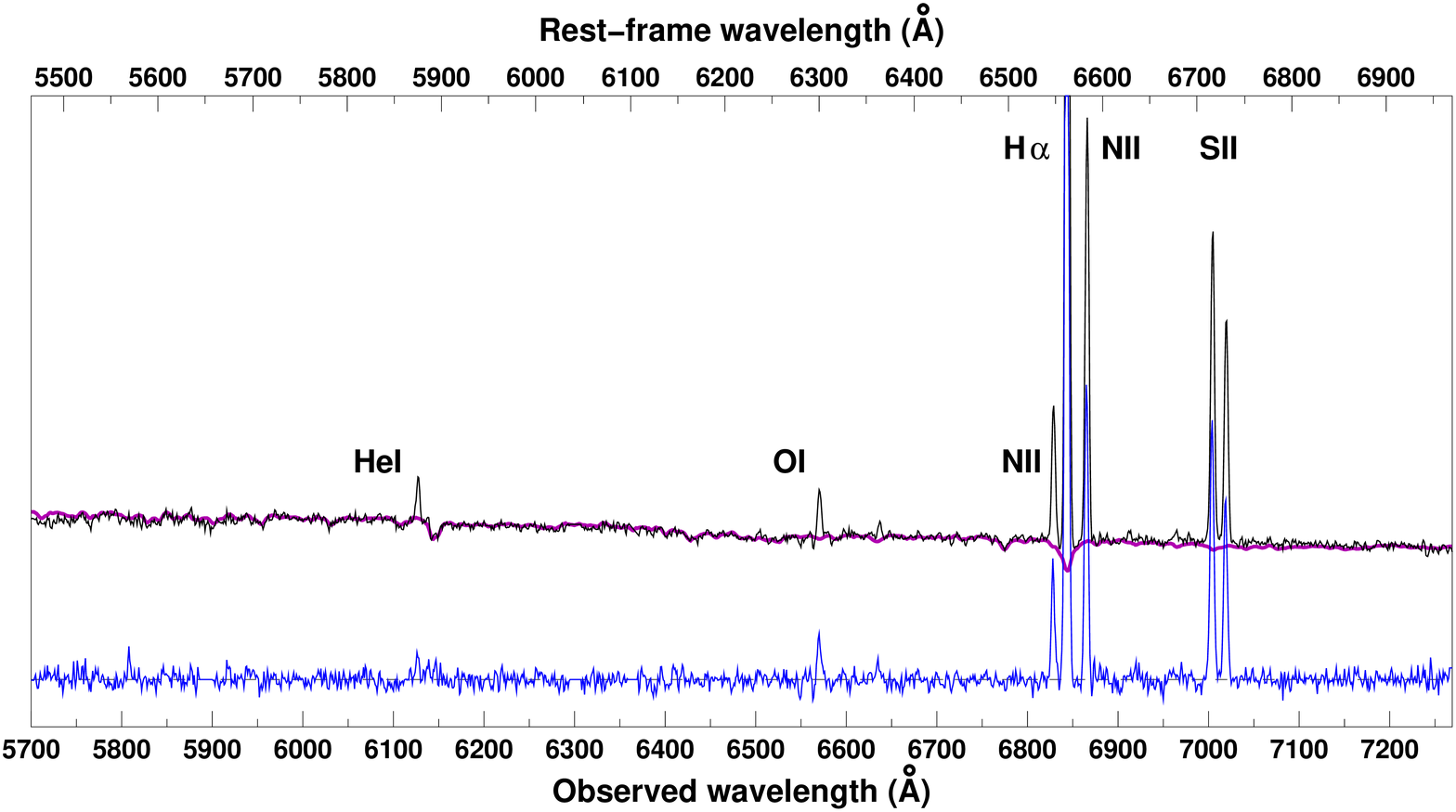}
\caption{{\bf SOS\,61086} Example results from the stellar continuum
  fitting process of the blue arm of SOS\,61086 (upper panel) and
  extrapolated to the red arm (lower panel). In the upper panel, the
  black curve shows the input spectrum, coming from the 3x3 spaxels
  region around the photometric centre of the galaxy. The magenta
  curve shows the resultant best-fit stellar continuum comprising a
  linear combination of SSPs, requiring both old ($>1$\,Gyr; thin red
  curve) and young ($<1$\,Gyr old; thin blue curve) components. The
  shaded regions indicate the wavelength ranges excluded from the
  fitting process, including the masks for emission lines. The
  residual emission component (thick blue curve) reveals clear
  emission at [O{\sc ii}]\,$\lambda 3729$, H$\zeta$, H$\epsilon$,
  H$\delta$, H$\gamma$, H$\beta$ and [O{\sc iii}]\,$\lambda
  \lambda\,4959\,5007$ in the blue arm (upper panel) and the emission
  lines of [O{\sc i}], H$\alpha$, [N{\sc ii}], [S{\sc ii}].}
\label{fspect_SOS61086}
\end{figure*}

\begin{figure*}
\includegraphics[width=164mm]{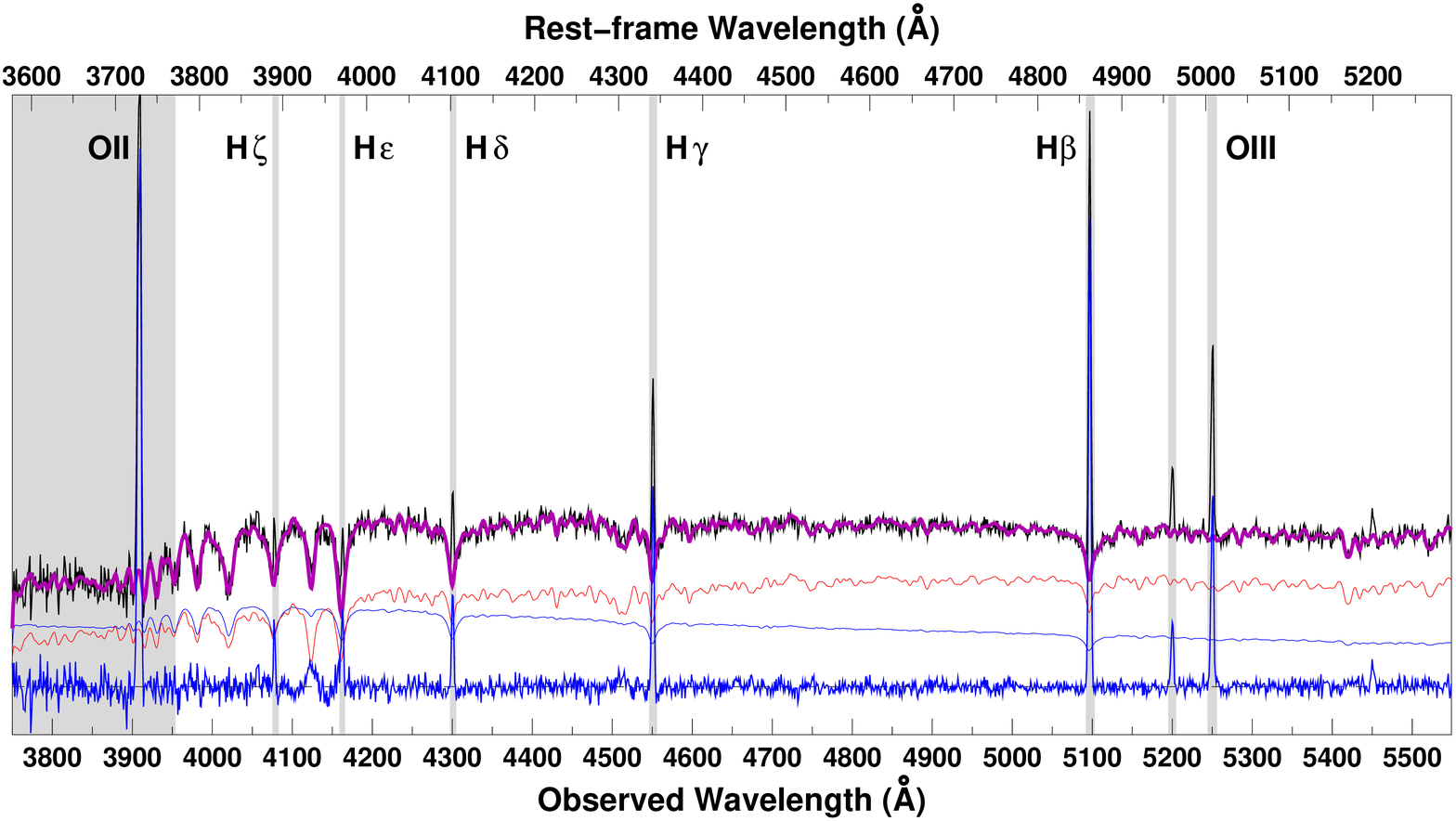}
\includegraphics[width=164mm]{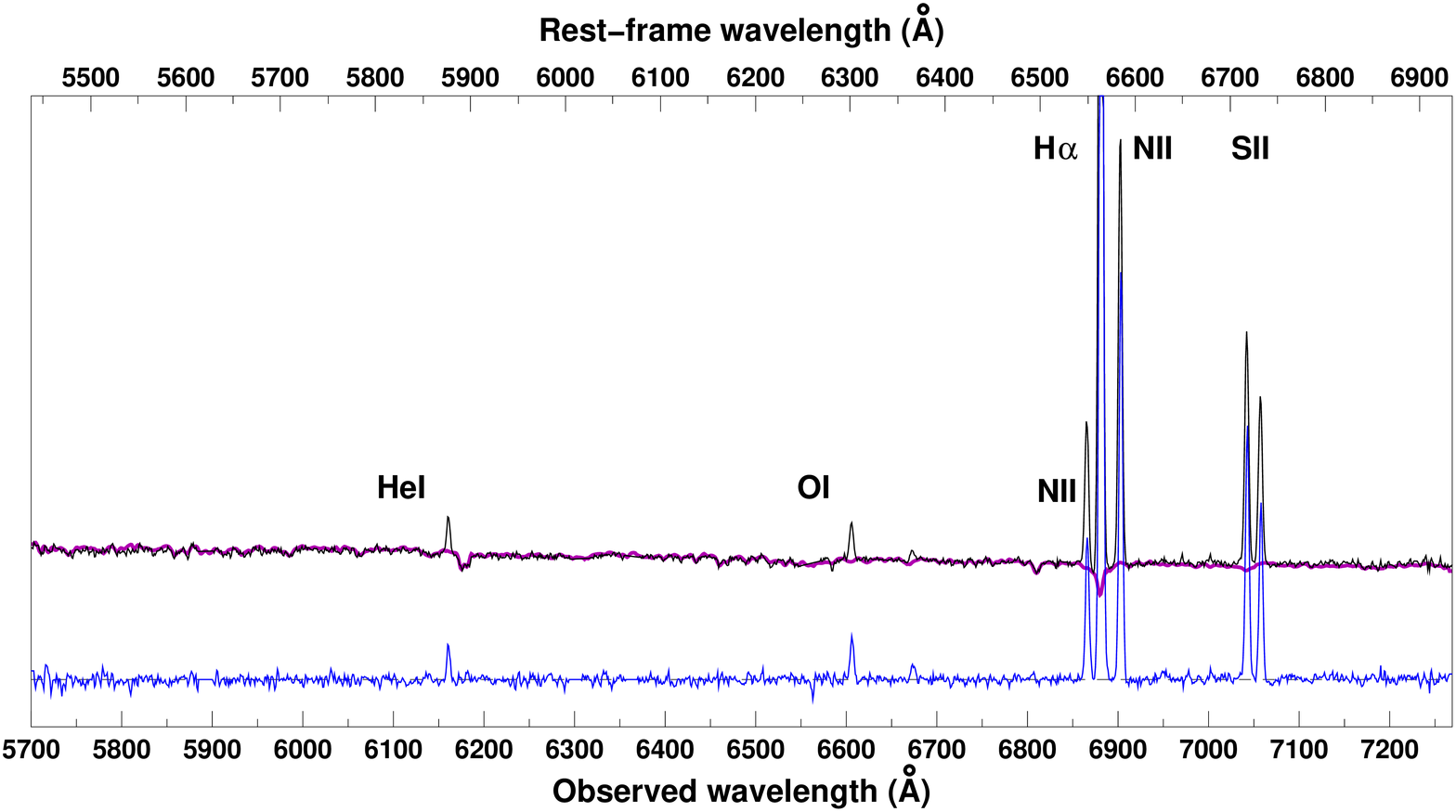}
\caption{{\bf 90630} Same as in Fig.~\ref{fspect_SOS61086}.}
\label{fspect_SOS90630}
\end{figure*}

\section{H$\bmath{\alpha}$ imaging}
\label{mmtf}

H$\alpha$ imaging of the galaxies SOS\,61086 and SOS\,90630 was
obtained with the Maryland-Magellan Tunable Filter
\citep[MMTF;][]{veilleux} on the Magellan-Baade 6.5m telescope at the
Las Campanas Observatory in Chile on 21 May 2012. The MMTF is based on
a Fabry-Perot etalon, which provides a very narrow transmission
bandpass (${\sim}$5--12{\AA}) that can be tuned to any wavelength over
${\sim}5$000--9200{\AA} \citep{veilleux}. Coupled with the exquisite
image quality provided by active optics on Magellan and the
Inamori-Magellan Areal Camera \& Spectrograph (IMACS), this instrument
is ideal for detecting extra-galactic H$\alpha$-emitting gas. The MMTF
6815-216 order-blocking filter with central wavelength of 6815{\AA}
and FWHM of 216{\AA} was used to provide coverage of the H$\alpha$
emission line for galaxies belonging to the Shapley supercluster
\citep[for details on the instrumental set-up see][]{ACCESSV}.
   
SOS\,90630 was observed for 75 minutes in H$\alpha$
($5{\times}900$\,s) and 30 minutes in the continuum, by shifting the
central wavelength of the etalon ${\sim}6$0{\AA} bluewards to exclude
emission from both the H$\alpha$ line and the adjacent [N{\sc ii}]
lines, and into a wavelength region devoid of major
skylines. SOS\,61086 was observed for 45 minutes in H$\alpha$
($3{\times}900$\,s) and 15 minutes in the continuum. The typical
image resolution for these exposures was $0\farcs60$.

These data were fully reduced using the MMTF data reduction
pipeline\footnote{http://www.astro.umd.edu/$\sim$veilleux/mmtf/datared.html},
which performs bias subtraction, flat fielding, sky-line removal,
cosmic-ray removal, astrometric calibration and stacking of multiple
exposures \citep[see][]{veilleux}. Photometric calibration was
performed by comparing the narrow-band fluxes from continuum-dominated
sources with their known $R$-band magnitudes obtained from our
existing SOS images. Conditions were photometric throughout and the
error associated with our absolute photometric calibration is
${\sim}10$\%. The effective bandpass of the Lorenzian profile of the
tunable filter of ${\pi}/2{\times}$FWHM is then used to convert the
observed measurements into H${\alpha}$ fluxes in units of
erg\,s$^{-1}$\,cm$^{-2}$. The $\Delta\lambda$ of the filter is
sufficiently narrow that there should be little or no contamination
from [N{\sc ii}] emission.

The data were obtained in dark time resulting in very low sky
background levels, with 1$\sigma$ surface brightness fluctuations
within a 1\,arcsec diameter aperture of
$0.2{\times}10^{-17}$\,erg\,s$^{-1}$\,cm$^{-2}${\AA}$^{-1}$\,arcsec$^{-2}$
implying a sensitivity of H${\alpha}$ imaging of
1.0$\times$10$^{-17}$\,erg\,s$^{-1}$\,cm$^{-2}$\,arcsec$^{-2}$ at
SNR=5, somewhat less than for the sensitivity of the spectroscopy.

\section{Data Analysis}
\label{DA}

In order to derive a robust estimate of the emission line fluxes the
contribution of the stellar continuum must be identified and
subtracted from the spectrum to leave the pure emission-line
spectrum. The stellar continuum modelling, accounting for the dust
extinction, also allows us to infer stellar population ages in
different galaxy regions. The emission-line fluxes are used to
estimate: i) the gas kinematics; ii) the line-ratio diagnostics; iii)
the dust attenuation and SFR across the galaxy. Details of this
analysis are given in \citet{ACCESSV}, here we briefly summarize the
different steps of the procedures.

\begin{figure}
 \begin{center}
\includegraphics[width=74mm]{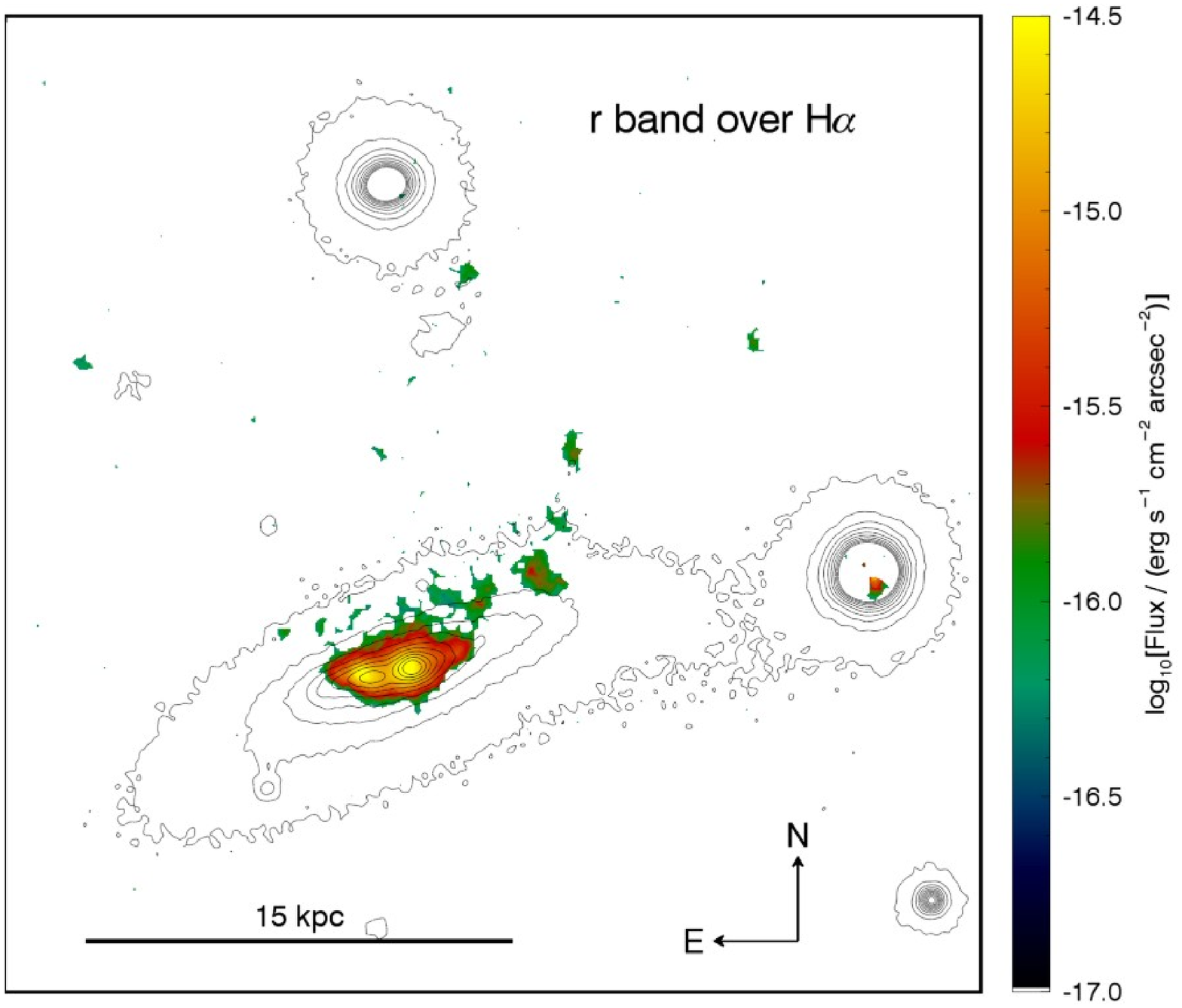}
\end{center}
\includegraphics[width=80mm]{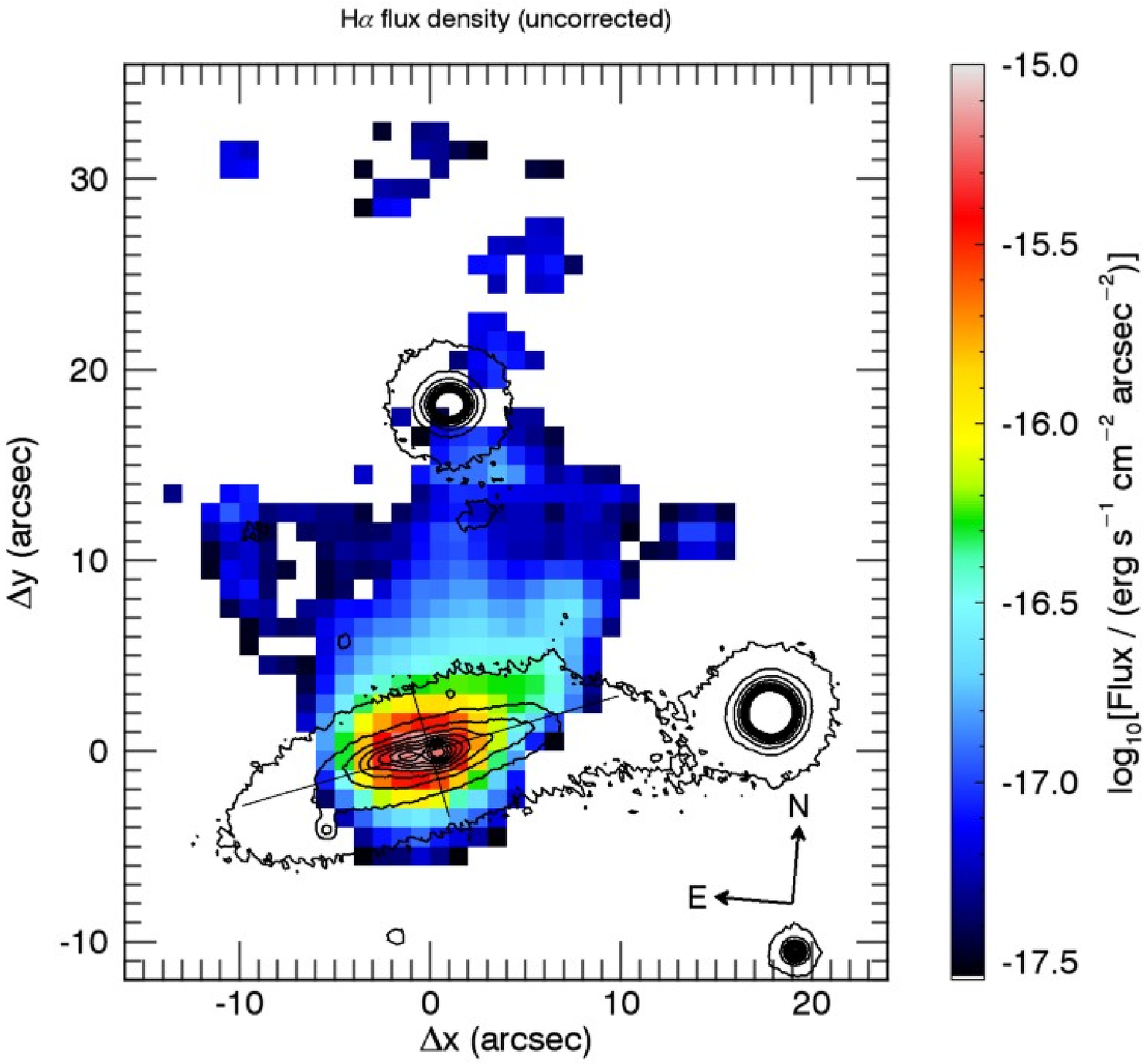}
\caption{{\bf SOS\,61086} Top: contours of the $r$ band over the MMTF
  H$\alpha$ image. Bottom: H$\alpha$ flux derived from IFS data. The
  $r$-band contours are shown in red and the H$\alpha$ flux in
  black. Here and in the following figures the scales give the
  distance (in arcsec) from the photometric centre of the galaxy. The
  major and minor axis are marked by lines of length three times the
  disk scale radius.}
\label{Ha61086}
\end{figure}

\subsection{Stellar continuum modeling and subtraction}
\label{anl2}

Late-type galaxies present complex star formation histories with
continuous bursts of star formation from the earliest epochs right up
until the present day \citep{K83,KTC94,JPB08,WDJ11}.  We thus attempt
to fit the stellar continuum of our target spiral galaxies as a linear
combination of 40 simple stellar populations (SSPs) from the
\citet{VSF10} stellar population models covering the full range of
stellar ages (0.06--15\,Gyr) and three different metallicities
[M/H]=--0.41, 0.0, +0.22. The models assume a \citet{K01} initial mass
function (IMF). They are based on the Medium resolution INT Library of
Empirical Spectra (MILES) of \citet{SBJ06}, have a nominal resolution
of 2.3\,{\AA}, close to our instrumental resolution, and cover the
spectral range 3540--7410\,{\AA}.

The spectra were smoothed spatially (using a $3{\times}3$ spaxel
region within the main galaxy body) to achieve a signal-to-noise ratio
of ${\sim}40$/{\AA} for the stellar continuum at
${\sim}$4600--4800\,{\AA}. For each spaxel, the spatially-smoothed
spectrum from the blue arm was fitted with the \citet{VSF10} models
after masking out regions below 3950\,{\AA}, which have significantly
reduced S/N levels and flux calibration reliability, above
5550\,{\AA}, where a bright sky line is located, and regions affected
by emission lines. The spectral pixels masked are kept fixed for all
spaxels, and for each emission line we carefully examine each datacube
to identify the full range of spectral pixels that are affected by the
emission/sky line for at least one spaxel, ensuring that the mask is
sufficiently generous to account for the shift of the emission line
due to velocity gradients across the galaxy (including the
extra-galactic gas). After continuum subtraction, we check the
datacube representing the residual emission-line spectrum to confirm
that all apparent emission lines (including very faint ones) have been
completely covered by the masks. The best-fit linear combination of
SSPs to the spatially-smoothed spectrum is then renormalized to fit
the spectrum from the individual spaxel (again excluding the masked
spectral pixels).  This process allows us to subtract the continuum
for spaxels in the outer regions of the galaxy where there is a clear
detection of the continuum, but the signal-to-noise ratio is much too
low to reliably fit complex stellar population models.

This best fit model was then extended to the red arm, and varying only
the global scaling factor to account for any (slight) mismatch in the
flux calibrations between red and blue arms, and subtracted from the
red arm spectrum, to produce the pure emission spectrum for the
corresponding spaxel in the red arm. Details on the stellar continuum
modelling are given in \citet{ACCESSV}.

In Figs.~\ref{fspect_SOS61086} and \ref{fspect_SOS90630} we show two
example outputs of our stellar continuum fitting process. The input
spectrum (black curve) is from a 3$\times$3 spaxels region in the
centre of the galaxy. The resultant best-fit stellar continuum
(magenta curve) comprising a linear combination of SSPs, requires both
young and old (thin blue/red curves) components. The shaded regions
correspond to wavelength ranges not considered in the fitting process,
including the masks for emission lines. This stellar continuum is
subtracted from the observed spectrum to produce the residual emission
component (thick blue curve) accounting now for stellar absorption,
revealing clear emission at O{\sc ii}, H$\zeta$, H$\epsilon$,
H$\delta$, H$\gamma$, H$\beta$ and [O{\sc iii}]. The Balmer emission
lines are all located within deep absorption features, demonstrating
the necessity of accurately modelling and subtracting the stellar
continuum prior to measuring these lines. Outside of the emission
lines, the rms levels in the residual signal are consistent with
expectations from photon noise, with little remaining structure. This
holds true throughout the galaxy indicating that on a spaxel-by-spaxel
basis the model fits to the stellar continuum are formally good
($\chi^{2}_{\nu}{\sim}1$).

The inverse problem of recovering the star formation history of
galaxies from their spectra is potentially ill-posed, and small
perturbations of the data due to noise can introduce large
perturbations in the best-fit stellar age distribution, at least in
terms of the relative contributions from individual SSPs. To modulate
this, many stellar population fitting codes employ a regularizing
method to minimize the curvature (burstiness) of the star formation
history while still being consistent with the observations. As our
primary objective is to accurately model the spectrum and subtract the
continuum so that it does not bias our emission line analyses, and map
the stellar kinematics, our code does not include a regularization
step or require the solution to be a relatively smooth function of
age. We confirm that the addition of random noise (at the levels of
the sky background) the relative contributions from individual SSPs do
vary significantly, but that this is mostly just shifting weights
among the SSPs that are adjacent in terms of stellar age (they are
separated by just 0.1\,dex). When consolidating the contributions into
a small number of age ranges, such as (above and below 1\,Gyr), the
perturbations due to noise on the best-fit model star formation
history are much less dramatic, barely shifting the relative
normalizations of the thin red and blue curves in Figure 8. We caution
the reader to consider our descriptions of the star formation history
obtained from this continuum fitting to be qualitative rather than
finely structured.

\subsection{Emission-line measurements}
\label{anl1}

Emission-line fluxes and widths were measured with Gaussian
fitting. Where lines are either partially overlapping or close (as in
the cases of [N{\sc ii}]-H$\alpha$-[N{\sc ii}], [S{\sc ii}]6717-6731
and [O{\sc iii}]4959-5007), the lines were fitted simultaneously but
independently to test our results against those line pairs having
fixed flux ratios. The uncertainties in the derived quantities were
evaluated through numerical simulations.

At SNR=5 the relative errors from the fit for both flux and $\sigma$
is $\sim$30\%. The total error on the flux also accounts for the
uncertainty in the flux zero-point, which amounts to $\sim$8\%.

The main contribution to the uncertainty on the radial velocity comes
from the uncertainty in the wavelength calibration
(${\sim}13$\,km\,s$^{-1}$), while the error from the fit is very small
(${<}5$\,km\,s$^{-1}$). We assume a conservative uncertainty of
$\Delta V{\sim}15$\,km\,s$^{-1}$.  The total error on the velocity
dispersion accounts for the error from the fit
(${\sim}1-10$\,km\,s$^{-1}$) and the uncertainty of
${\sim}5$\,km\,s$^{-1}$ on the value of the instrumental width.

\begin{figure*}
\includegraphics[width=160mm]{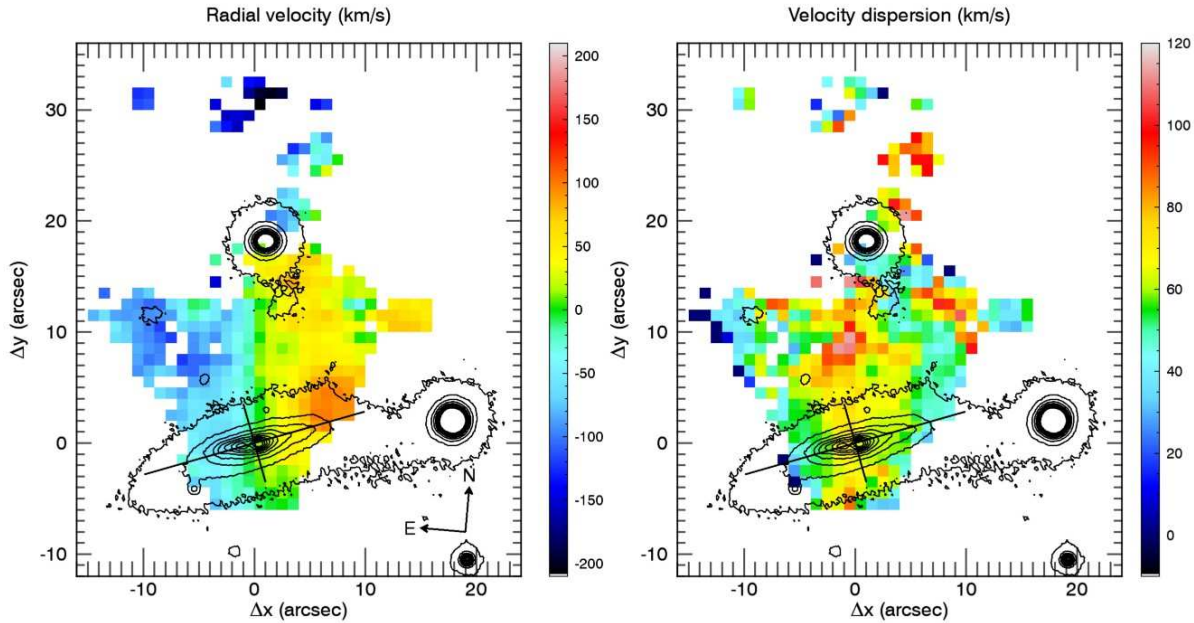}
\caption{{\bf SOS\,61086} Gas kinematics derived from the fit to the
  H$\alpha$ emission line. The black contours mark the surface
  brightness distribution of the $r$-band image, while the black
  lines, extending 3$\times$ the disk scale radius, show the positions
  of the minor and major axis and cross at the $K$-band photometric
  centre.  {\it Left}: Gas velocity field. Each pixel is colour coded
  according to the measured radial velocity relative to the galaxy
  centre. {\it Right}: Gas velocity dispersion. The
  velocity scales are on the right of each panel.}
\label{SOS61086kin}
\end{figure*}

In \citet{ACCESSV} we have shown that, in order to obtain flux ratios
with uncertainties lower than 30\%, the individual lines must have
SNR$>$20. To achieve these SNRs we spatially binned the data by means
of the \emph{`Weighted Voronoi Tessellation'} (WVT) method by Diehl \&
Statler\footnote{http://www.phy.ohiou.edu/diehl/WVT} (\citeyear{DS06})
which attempts to reach a target SNR in all bins. The WVT performs a
partitioning of a region based on a set of points called ‘generators’,
around which the partition of the plane takes place. The partitioning
is repeated iteratively until a target SNR is achieved in all
bins. The advantage of the WVT with respect to the Voronoi
tessellation is the possibility to ‘manually’ choose a number of
generators on the basis of their position within the galaxy. The
faintest line involved in our flux ratios is the [O{\sc
    i}]\,$\lambda$6300 line, which is $\sim 5\times$ fainter than
H$\alpha$, so that we set to 100 the target SNR of H$\alpha$.  For
spatial consistency, we manually adjusted some of the WVT regions. The
35 and 48 galaxy regions identified by the WVT for SOS\,61086 and
SOS\,90630 respectively are shown in Figs.~\ref{SOS61086bins} and
\ref{SOS90630bins} (see Sects.~\ref{SF_1} and \ref{SF_2},
respectively). The spatial binning of the data was applied for the
derivation of flux ratios, dust attenuation, and SFR, but not for the
kinematics, for which no binning was necessary.

\section{SOS\,61086: results}
\label{RES61086}

\subsection{Morphology of the H$\alpha$ emission}
\label{Halpha_1}

The narrow-band H$\alpha$ image and the distribution of H$\alpha$
emission derived with WiFeS are shown in Fig.~\ref{Ha61086} (upper and
lower panel respectively). Since the H$\alpha$ imaging is both sharper
and shallower than the WiFeS data, the two data sets are complementary
and together show different aspects of the distribution of ionised
gas. The IFS data show that ionised gas spreads out from the disk of
the galaxy in an approximately triangular region with a vertex in the
central disk and one side at $\sim$16\,kpc North and directed
approximately E-W. No ionised gas is found in the disk beyond
$\sim$6\,kpc from the centre along the major axis. Along the minor
axis, instead, the ionised gas appears to extend far out from the disk
in projection. Other clumps of gas extend further in the North
reaching $\sim$30\,kpc in projection. The H$\alpha$ flux is maximum in
the centre and is directed along a main stream in the NW
direction. Three secondary maxima of flux are found along the northern
side. The narrow-band image resolves the nuclear emission in two main
clumps lying around the photometric centre as determined by the
maximum in K-band flux. The NW stream is also resolved in a chain of
major emission clumps departing from the centre. Other H$\alpha$
clumps are also detected in N and NE directions. The nearby galaxy
SOS\,61087 does not show any H$\alpha$ emission associated with it.

\subsection{Gas and stellar kinematics}
\label{GSK_1}

The radial velocity field of the gas derived from H$\alpha$ is shown
in the left panel of Fig.~\ref{SOS61086kin}. The black contours trace
the $r$-band stellar continuum derived from VST imaging. The black
lines mark the major and minor axes of the disk extending to 3$\times$
the disk scale radius, and crossing at the $K$-band photometric
centre. The kinematic centre coincides with the photometric centre of
the $K$-band image.

Given the strong perturbation experienced by the gas, the most
remarkable feature of this velocity field is the large-scale regular
motion. Overall, the gas appears to rotate around an axis extending
$\sim$12\,kpc (in projection) North from the galaxy centre, well
beyond the stellar disk. This `rotation axis' bends westwards South
from the centre. Superimposed to the large-scale motion there are
however some significant local maxima and minima in radial velocity,
like for instance an area at $\sim$7\,kpc W from the centre and
$\sim$1\,kpc N from the disk, with V$\sim$85\,km\,s$^{-1}$. The
distant shreds of gas in the North do not seem to participate in this
overall motion, being generally blue-shifted with respect to the
average.

The gas velocity dispersion $\sigma$ (Fig.~\ref{SOS61086kin} right
panel) is generally higher than expected for normal turbulent
\HII~regions ($\sigma{\sim}2$0--30\,km\,s$^{-1}$).  As expected,
$\sigma$ is higher where the radial velocity gradients are larger.
But in some areas it reaches values as high as 100\,km\,s$^{-1}$, such
as in the region about 8\,kpc N from the centre.  Such high values
most probably witness the presence of complex motions of different gas
elements along the line of sight.

\begin{figure}
\begin{center}
\includegraphics[width=80mm]{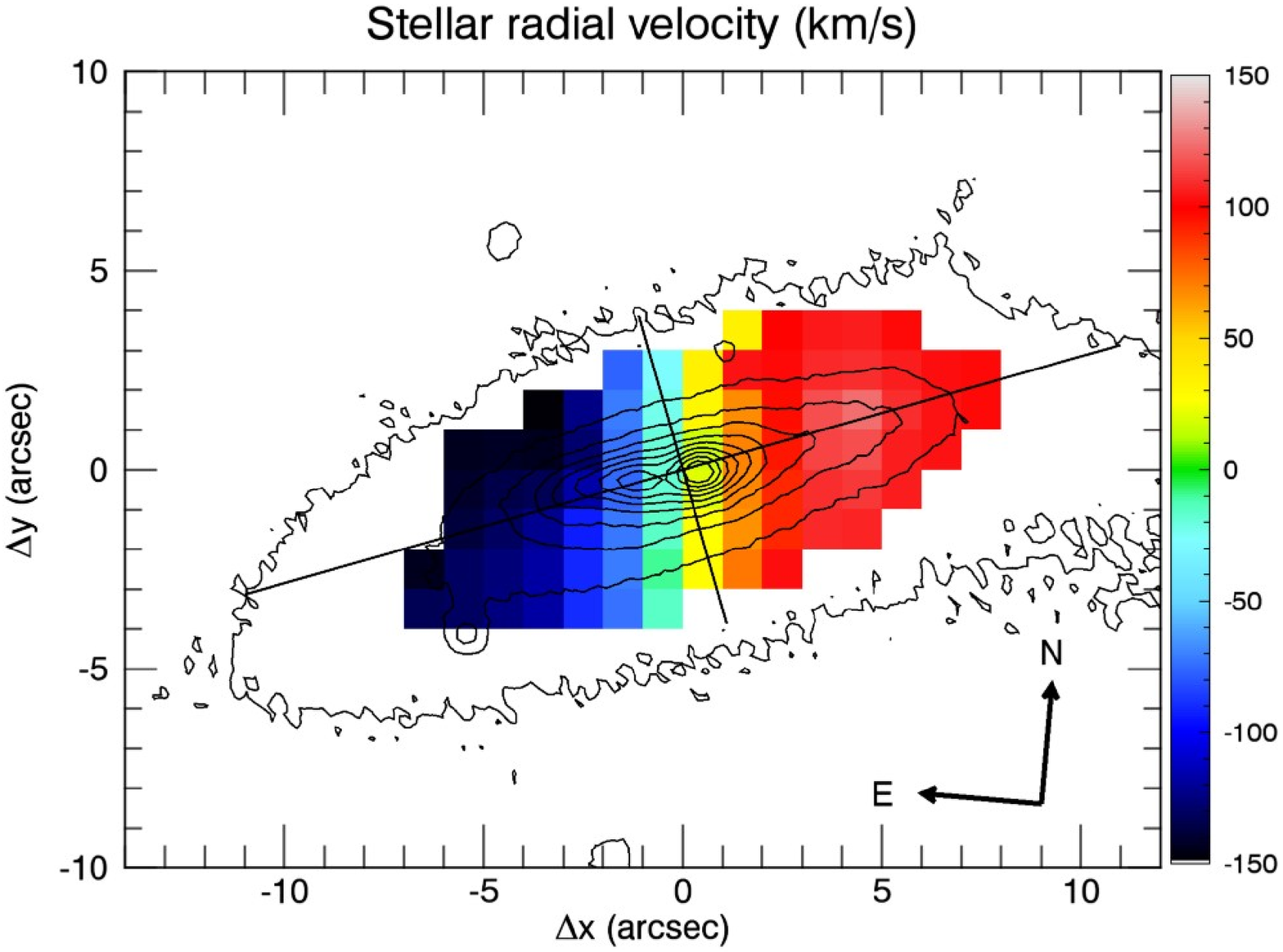}
\end{center}
\includegraphics[width=80mm]{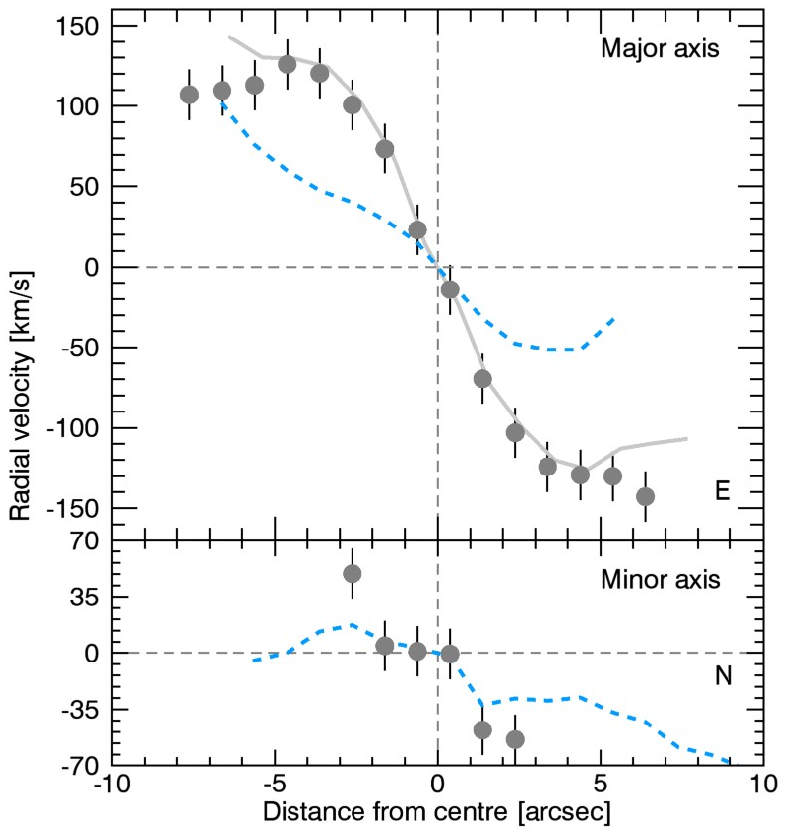}
\caption{{\bf SOS\,61086} The stellar velocity field (top). Same symbols and
  $r$-band contours as in Fig.~\ref{SOS61086kin}. Radial velocity
  profiles along the major and minor axis (bottom) of stars (grey
  dots) and gas (blue dashed curve).  The grey curve is the flipped
  stellar velocity profile, plotted to assess its symmetry.  Errors of
  15\,km\,s$^{-1}$, mainly coming from wavelength calibration, are
  indicated.}
\label{SOS61086sk}
\end{figure}

From the fit of the stellar continuum we derived the velocity field of
the stars shown in Fig.~\ref{SOS61086sk}.  The stellar velocity field
is fairly symmetric in the inner $\sim$4\,kpc. The kinematic axis is
skewed with respect to the minor axis, reflecting the presence of
non-circular motions, most probably to the presence of a bar. This
axis is aligned with the kinematic axis of the gas.  Also the
the shape of the velocity field bears some resemblance with the gas, but the
comparison of the two velocity profiles (lower panel of
Fig.~\ref{SOS61086sk}) show that the two components are clearly
decoupled. The shape of the minor-axis velocity profile of the stars
is consistent with the non-circular motions due to a bar.

As a note of caution, we must remark that the presence of dust adds
complications to the interpretation of the kinematics, since dust may
selectively hide parts of the moving gas or stars
\citep[e.g.][]{GH02,BDD03,KVF04,VG04}.

\subsection{Dust extinction across the galaxy}
\label{SF_1}
\begin{figure}
\begin{centering}
\includegraphics[width=60mm]{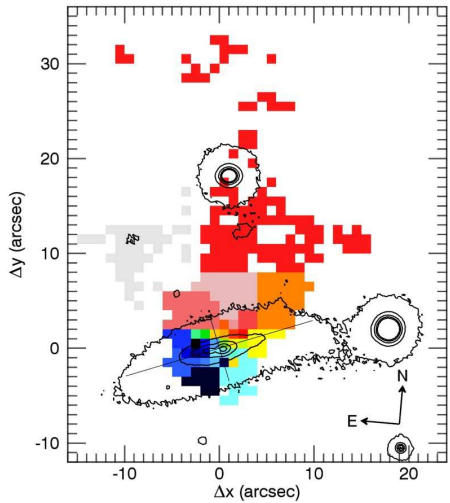}
\caption{{\bf SOS\,61086} The 35 galaxy regions identified by the WVT
  are shown in different colours. For each of these regions reliable
  and robust measurements of emission line ratios are derived. The
  colour coding shown here is used to identify the region
  corresponding to each symbol in Fig.~\ref{SOS61086KBPT}.}
\label{SOS61086bins}
\end{centering}
\end{figure}

For the subsequent analysis of this galaxy, we binned the spatial
pixels to achieve the needed SNR as explained in
Sect.~\ref{anl1}. Only the pixels with SNR(H$\alpha$)$>8$ were
considered to make the bins\footnote{In spaxels with
  SNR(H$\alpha$)$\sim$3, many fainter lines (notably [O{\sc
      i}]$\lambda$6300) are not even visible. We therefore adopted a
  higher threshold for SNR(H$\alpha$), with a value around 8
  representing a good compromise between quality and extension of the
  measured area.}. The 35 regions so identified are shown in
Fig.~\ref{SOS61086bins}.

\begin{figure}
\begin{centering}
\includegraphics[width=80mm]{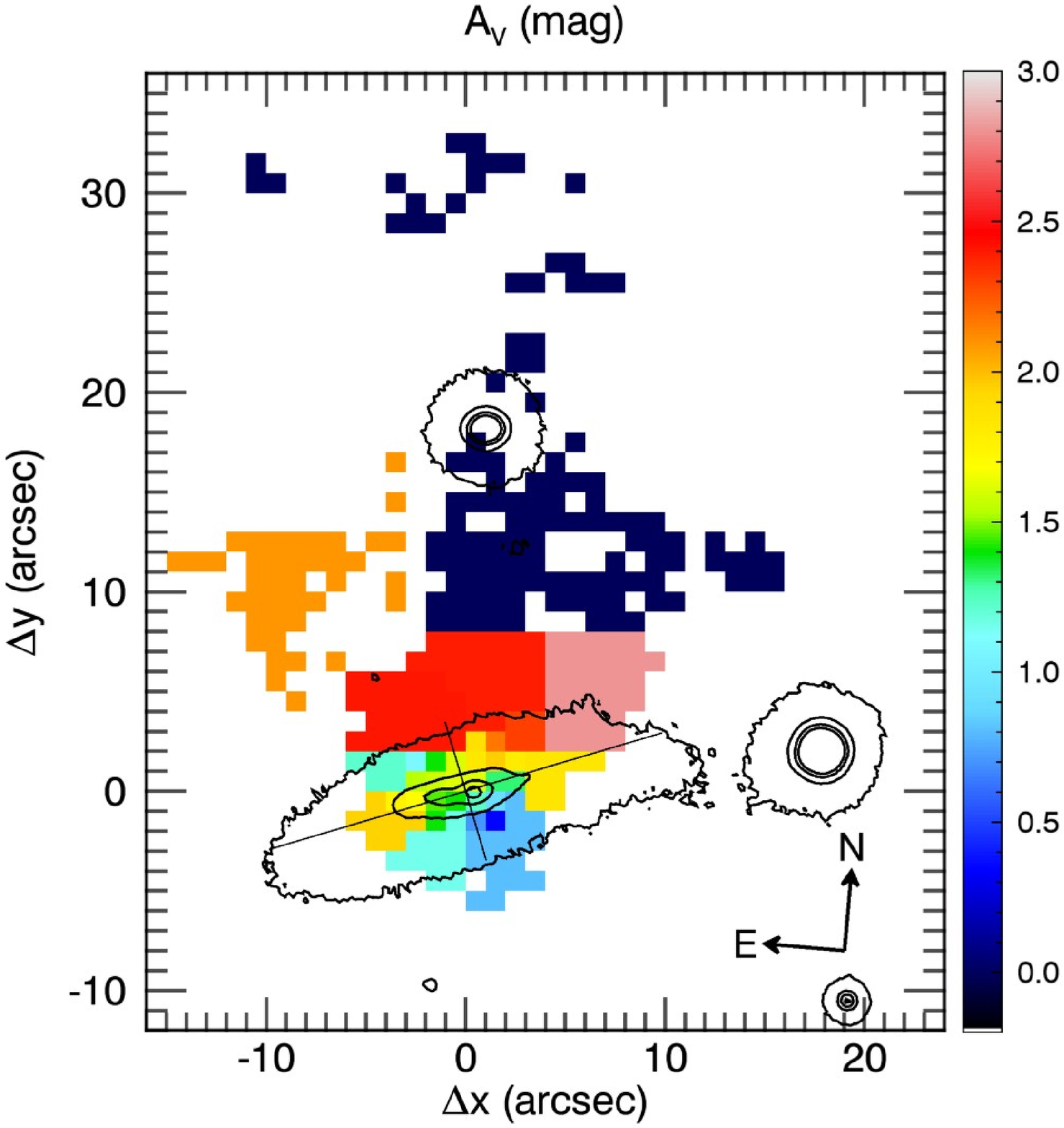}
\includegraphics[width=80mm]{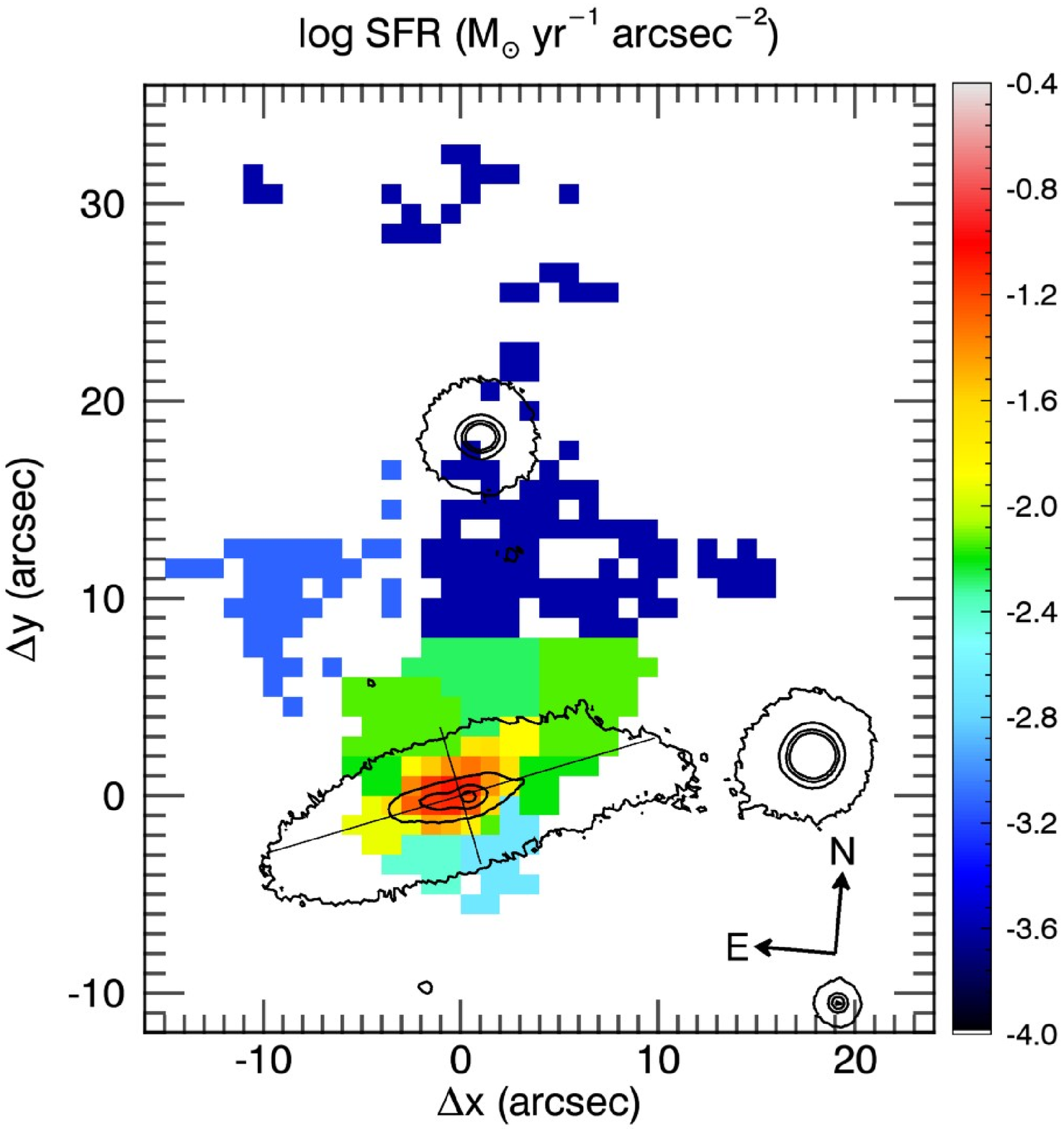}
\end{centering}
\caption{{\bf SOS\,61086} Dust attenuation derived from the
  H$\alpha$/H$\beta$ line ratio (top) and attenuation-corrected SFR
  (bottom). The contribution to the SFR from the NE
    (shock-dominated) tail is not considered in
    the analysis.}
\label{SOS61086SF}
\end{figure}

\begin{figure}
\includegraphics[width=85mm]{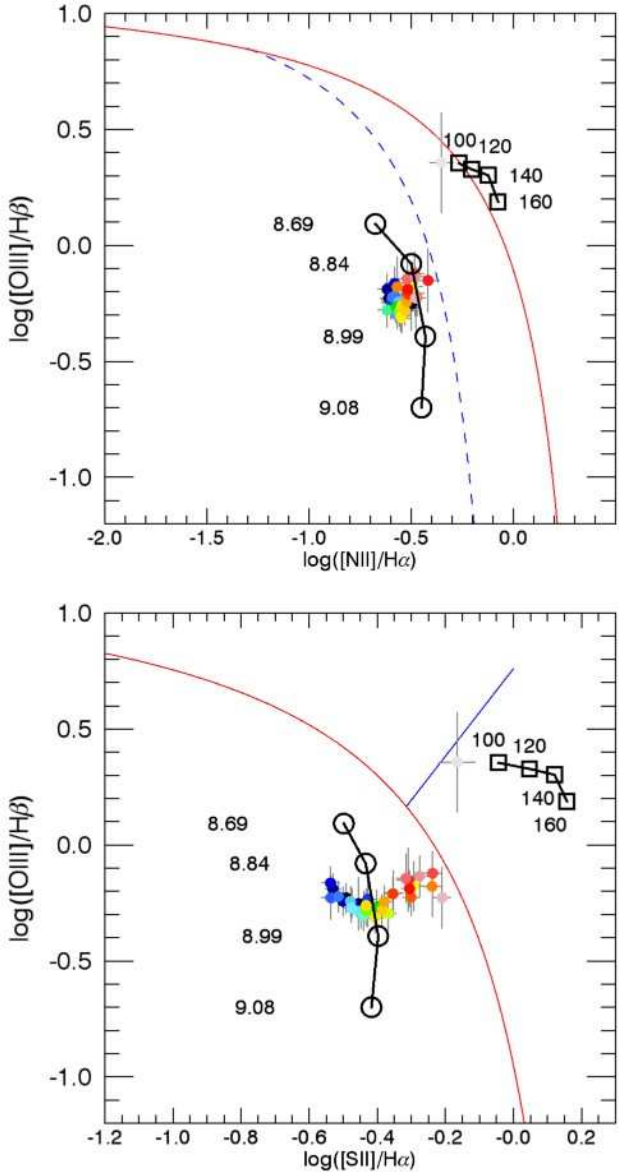}
\caption{{\bf SOS\,61086} Line flux diagnostic diagrams of the
  different galaxy regions (see text) colour coded as in
  Fig.~\ref{SOS61086bins}. The theoretical (red curve) and empirical
  (blue dashed curve in the upper panel) upper limits for \HII~regions
  are indicated as well as the separation between AGN and LINER (solid
  blue line in the lower panel). We also show a set of \HII\ region
  models for fixed ionisation parameter $\log q = 7.0$ and four
  abundances (open circles), as indicated, along with four shock
  models with shock velocities in the range $100-160$\,km~s$^{-1}$
  (open squares).}
\label{SOS61086KBPT}
\end{figure}

The distribution of the dust attenuation derived for each of these
regions from the H$\alpha$/H$\beta$ line ratio, given in terms of the
$V$-band extinction $A_V$ is shown on the upper panel of
Fig.~\ref{SOS61086SF}. Here we adopted an intrinsic ratio of 2.87 (the
`case B' recombination). In doing so, we implicitly assume that the
extraplanar gas is in the conditions of typical H{\sc ii} regions, in
which the primary source of ionisation is ongoing star
formation. Other sources of ionisation such as thermal conduction,
shock-heating or magneto-hydrodinamic processes
\citep[e.g.][]{VBC99,BCF16} might be at work in the tail, in which
case the actual $A_V$ would be significantly different from the one
estimated here. This is certainly the case for the NE region of the
tail (light gray in Fig.~\ref{SOS61086KBPT}), which we show below to
be dominated by shock heating. For the rest of the tail, the continuum
emission detected in the whole UV-optical wavelength range ($u$ to
$i$), the presence of compact emission regions witnessed by the
H$\alpha$ image and the diagnostic diagrams (see below) make us
confident that most of the detected emission is associated to star
formation \citep[e.g.][]{YYK12}. We adopted the theoretical
attenuation curve by \citet{FD05} with $R_{V}{=}4.5$. The highest dust
extinction ($A_V{\sim}2$.5-2.9\,mag) is observed in two regions
covering the NW disk and the extraplanar gas just out of the disk. The
asymmetric distribution of the dust extinction strengthens our
hypothesis on the apparent bending of the disk
(Sect.~\ref{SOS61086des}). Moderate ($A_V{\sim}1$.5\,mag) dust
attenuation is observed in the galaxy centre, while the southern part
of the disk and the NW tail of the extraplanar gas suggest a lower
dust content.

\subsection{Ongoing star formation and recent star formation history}

We derive the current SFR from the H$\alpha$ flux taking into account
the effects of dust extinction following \citet{K98} and adopting the
Kroupa IMF. The SFR surface density across the galaxy is given in the
bottom panel of Fig.~\ref{SOS61086SF}. The integrated
H$\alpha$-derived SFR of SOS\,61086 amounts to $1.76\pm
0.56$\,M$_\odot$yr$^{-1}$ where the error takes into account the
uncertainties related to the flux and attenuation measurements added
to a 30 per cent uncertainty due to the different calibrations of
H$\alpha$ as a SFR indicator \citep{K98}. The contribution to the SFR
from the NE (shock-dominated, see Sect.~\ref{BTP_1}) tail is not
considered. Half of the SF occurs in a central region of 2.5\,kpc
radius. Lower SFR is detected in the external disk and the extraplanar
gas close to the disk. The SFR value in the gas tail far from the disk
suggests that SF in low in the detached gas. The integrated
H$\alpha$-derived SFR is consistent with that derived from UV and IR
data (see Table~\ref{gals}).

\begin{figure}
\begin{centering}
\includegraphics[width=80mm]{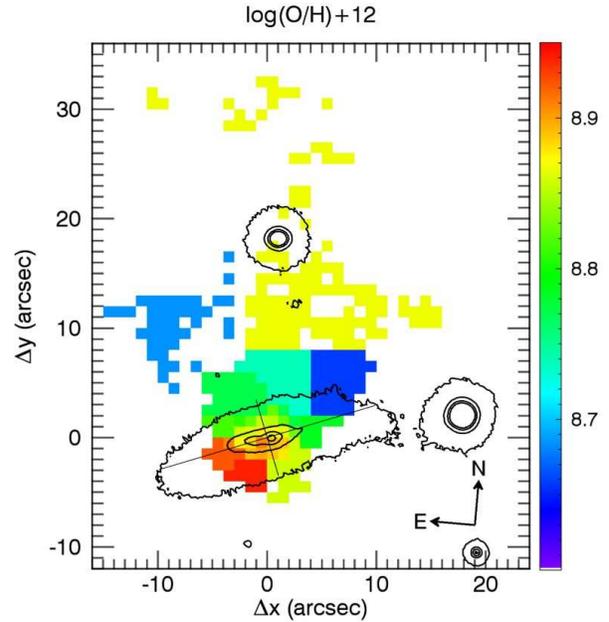}
\caption{{\bf SOS\,61086} Abundance distribution in the galaxy as
  derived using the {\tt Pyqz} module.}
\label{SOS61086abu}
\end{centering}
\end{figure}

The recent star formation history across the disk is constrained by
the stellar continuum modelling described in Sect.~\ref{anl2}. In the
centre of SOS\,61086, where the density of star formation appears
highest (see Figs.~\ref{fspect_SOS61086} upper panel), the emission
from the stellar continuum appears relatively blue, peaking at
${\sim}4$\,000{\AA} (rest-frame). This reflects the dominant young
stellar population in the centre of the galaxy, with stars younger
than 1\,Gyr (thin blue curve in Fig.~\ref{fspect_SOS61086} upper panel)
contributing 85\% of the luminosity at 4000{\AA}, and 24\% of the
stellar mass. This dominance of young stars is supported by the deep
H$\delta$ and Ca{\sc ii}\,H absorption features, which give rise to
low values of the Ca{\sc ii} index \citep[0.65;][defined as the ratio
  of the counts in the bottom of the Ca{\sc ii}\,H+H$\epsilon$ and
  Ca{\sc ii} K lines]{R85} and the H$\delta/Fe${\sc i}$\lambda$4045
index (0.53), indicative of an ongoing/recently terminated star-burst
that dominates the integrated light at 4000{\AA} \citep{LR96}. Our
stellar modelling of the young component suggests a continuous steady
SFR over the full range 60--1000\,Myr, with 16\% of the stellar mass
within this young component coming from stars younger than
200\,Myr. This sustained ${\sim}2{\times}$ increase in star formation
is still ongoing as implied by the strong H$\alpha$ emission in this
region with EW$({\rm H}\alpha){\sim}60${\AA} (see
Fig.~\ref{fspect_SOS61086} lower panel). Further Balmer emission lines
are visible in Fig.~\ref{fspect_SOS61086}, from H$\beta$ all the way
up to H$\zeta$ (3889{\AA}). The old stellar component (red curve)
includes both intermediate age (2--3\,Gyr old) and primordial
(8--15\,Gyr old) components, indicative of a continual formation of
stars through the lifetime of the galaxy.

As we move from the centre of the galaxy to the eastern edge the
spectra progressively change from ongoing starburst to post-starburst
signatures, with EW(H$\alpha$) declining steadily to zero, and the
H$\delta$ and Ca{\sc ii}\,H absorption lines becoming ever
deeper. Here the Ca{\sc ii} index \citep{R85} falls to a value of just
0.27, while the H$\delta/Fe${\sc i}$\lambda$4045 index also drops to
0.42. Such extreme values are too low for any of the burst models of
\citet{LR96}, but certainly require the entire integrated light at
4000{\AA} to be due to a recent/ongoing burst. Indeed, our best-fit
stellar population model has little (if any) contribution from old
($>$1\,Gyr) stellar populations (red curve) in terms of emission,
being dominated by 0.3--1.2\,Gyr stellar populations with spectra
dominated by A-type stars.

\subsection{Excitation and metallicity of the gas}
\label{BTP_1}
The line ratios and velocity dispersions observed in the ionised gas
enable us to distinguish between gas which is photoionised by
\HII\ regions, and gas which is excited by shocks. In order to
interpret the line ratio diagnostics, we have run some representative
shock and photoionisation models using the modelling code {\tt
  Mappings 4} described in \citet{Dopita13}. In Fig~\ref{SOS61086KBPT}
we show the observations of SOS\,61086 on the standard BPT diagrams
\citep{Baldwin81} as refined by \citet{VO87}, \citet{KHD01},
\citet{KHT03} and \citet{KGK06}. We also plot four \HII\, region
models generated as described in \citet{Dopita13} with a ionisation
parameter $\log q = 7.0$ (see below), and four different metallicities
(open circles in Fig.~\ref{SOS61086KBPT}). We also show, for
comparison, four shock models with 2$\times$ solar metallicity ($12 +
\log{\mathrm O/H} \sim 8.99$) and four different velocities (open
squares in Fig.~\ref{SOS61086KBPT}). These shock models have fully
self-consistent ionisation.

\begin{figure}
\begin{center}
\includegraphics[width=55mm,trim=0mm 0mm 3mm 0mm,clip=true]{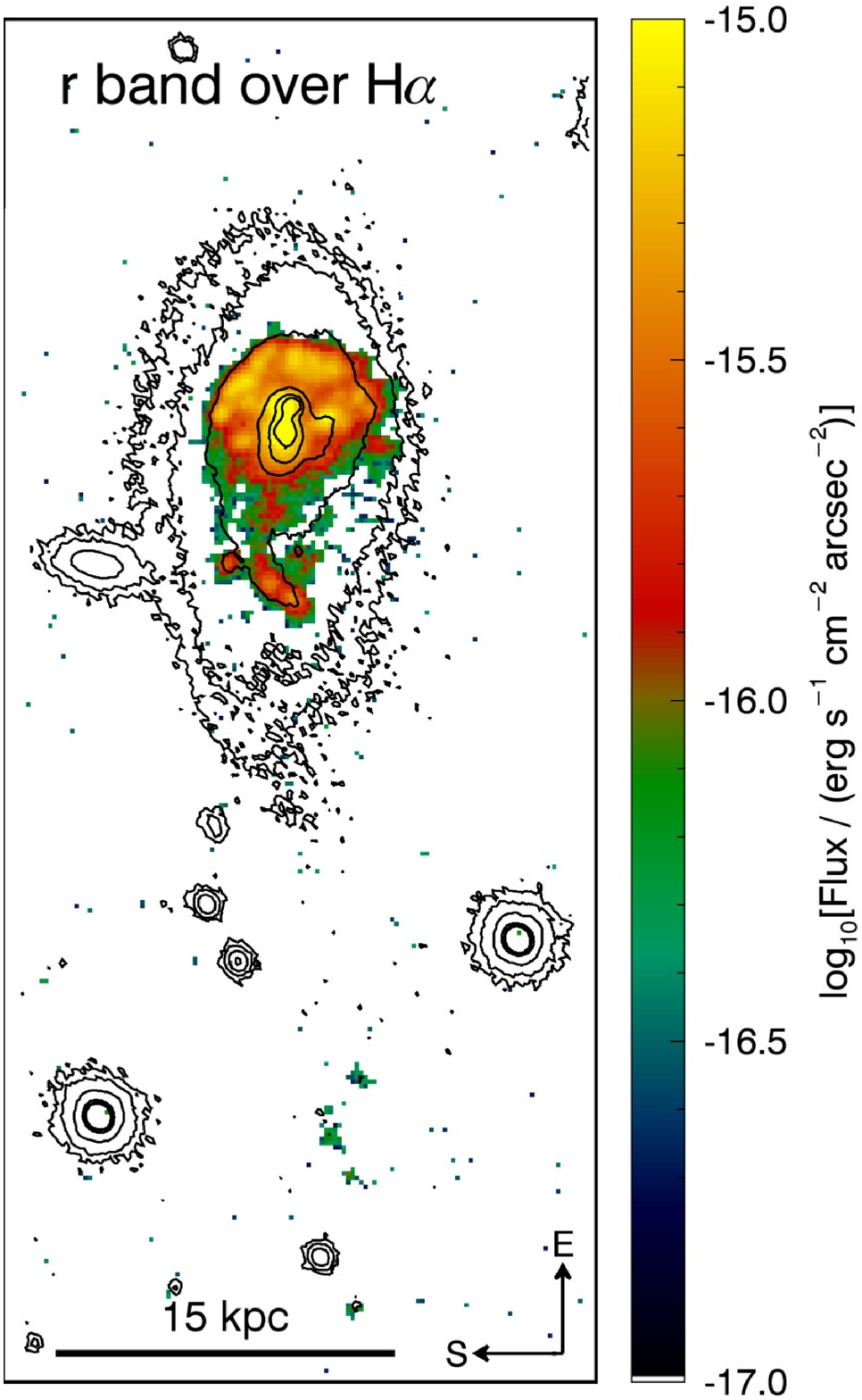}
\includegraphics[width=63mm,trim=3mm 0mm 0mm 0mm,clip=true]{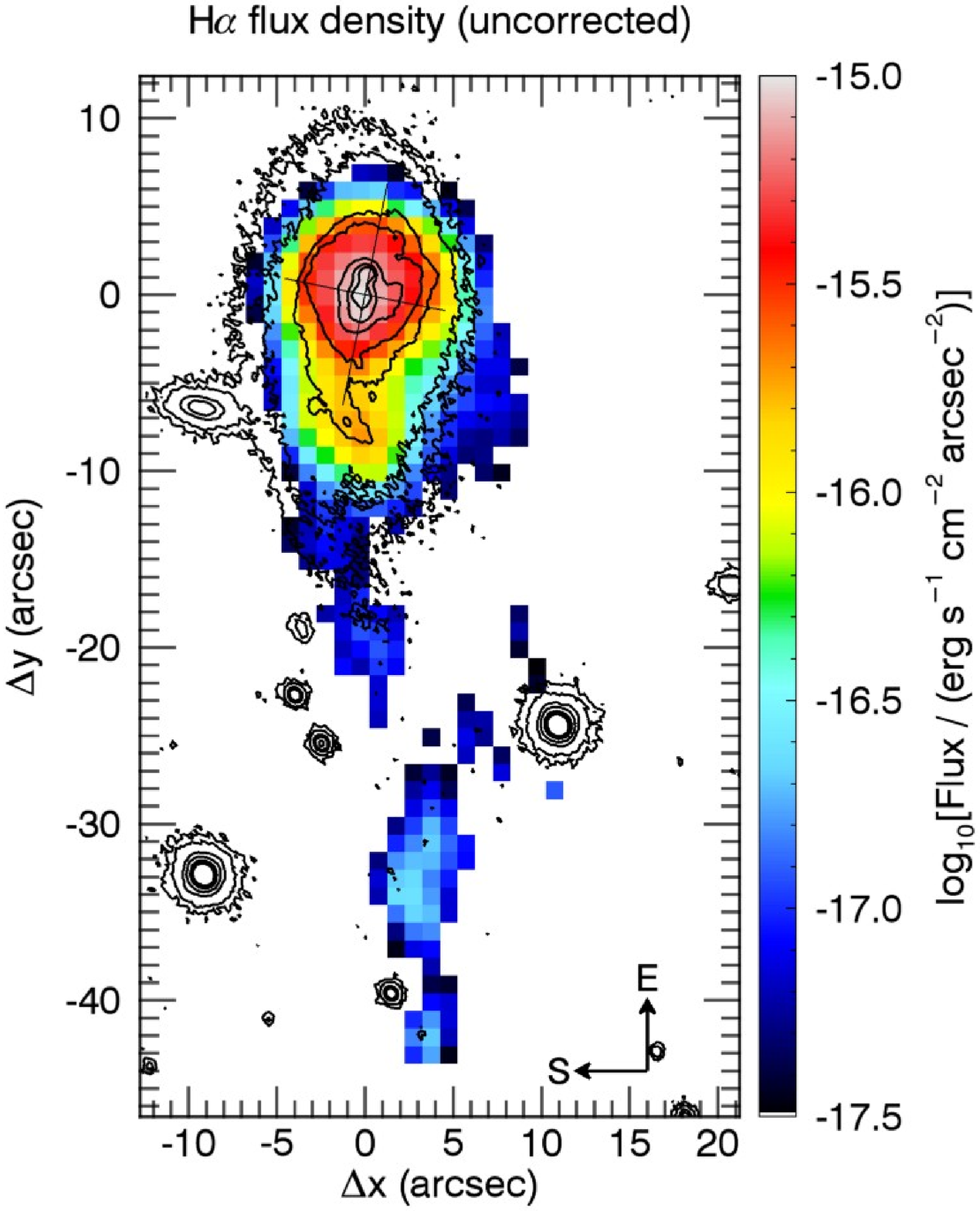}
\end{center}
\caption{{\bf SOS\,90630} Top: contours of the $r$ band over the MMTF
  H$\alpha$ image. Bottom: H$\alpha$ flux derived from IFS data. The
  $r$-band contours are shown in red and the H$\alpha$ flux in black.}
\label{Ha90630}
\end{figure}

In our first study \citep{ACCESSV} and in other cases of RPS galaxies
\citep[e.g.][]{FFB16} clear evidence has been found that both photo-
and shock-ionisation were at work. In SOS\,61086 the situation with
regard to shocks is different. The line ratios cluster over a very
narrow range (see Fig.~\ref{SOS61086KBPT}) consistent (according to
our photoionisation models) with gas of somewhat higher than solar
metallicity ($12 + \log{\mathrm O/H} \sim 8.8$) being ionised mainly
by hot stars, except for the NE tail. By comparison of the
observations with the shock sequence it is clear that shocks are
important in determining the excitation only in the NE region of the
tail.

The tight grouping of points indicates very little sign of an
abundance gradient in this galaxy, or in the extraplanar gas
associated with it. We have used the Python module {\tt
  Pyqz}\footnote{The lines in the {\tt Pyqz} code are H$\beta$, [O{\sc
      iii}]\,$\lambda\,5007$, [O{\sc i}]\,$\lambda\,6300$, H$\alpha$,
  [N{\sc ii}]\,$\lambda\,6583$, [S{\sc ii}]\,$\lambda
  \lambda\,6717-6731$. } described in \citet{Dopita13} to estimate the
ionisation parameter and the oxygen abundance in each of the 35 galaxy
regions identified by the WVT. The ionisation parameter spans a narrow
($\sigma_{\log{q}} = 0.08$) range of values around $\log{q}=
6.95$. The oxygen abundance is shown in Fig.~\ref{SOS61086abu} where a
rather uniform distribution is seen.

\section{SOS\,90630: results}
\label{RES90630}

\subsection{Morphology of the H$\alpha$ emission}
\label{Halpha_2}
The H$\alpha$ image and the distribution of H$\alpha$ emission from
IFS for SOS\,90630 are shown in Fig.~\ref{Ha90630} (upper and lower
panels respectively). The ionised gas disk appears truncated in the
ESE side.  Gas extending $\sim$4\,kpc in projection out of the disk is
seen in the NW side, and more prominently along a tail extending by
more than 41\,kpc to the West. The maximum of H$\alpha$ emission takes
place in the centre of the galaxy, and is resolved into two clumps
about 1\,kpc apart in the narrow-band image. The MMTF image resolves
an arc-shaped crown composed of H$\alpha$ emitting knots all long the
leading edge of the galaxy. Other knots of H$\alpha$ flux are detected
along a western arm and in the tail. The MMTF imaging also reveals
compact (${<}0.5$\,kpc) H$\alpha$ emission at the centre of
SOS\,90090, indicating the presence of an AGN or ongoing nuclear
star-formation activity. The overall morphology of the H$\alpha$
emission is a remarkable example of the classic ``jellyfish'' forms
reproduced in simulations of galaxies undergoing RPS
\citep[e.g.][]{KSS09}.

\begin{figure*}
\begin{centering}
\includegraphics[width=140mm]{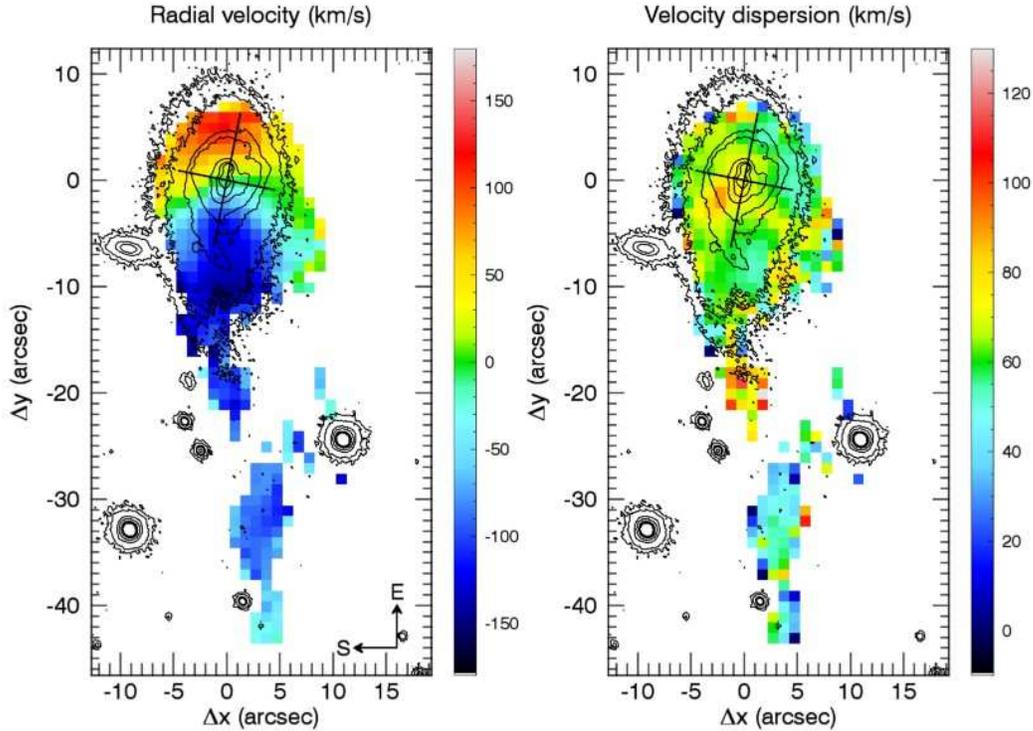}
\end{centering}
\caption{{\bf SOS\,90630} Gas velocity fields derived from the fit to
  the H$\alpha$ emission line. {\it Left}: radial velocity field. {\it
    Right}: velocity dispersion. Symbols as in Fig.~\ref{SOS61086kin}}.
\label{SOS90630kin}
\end{figure*}

\subsection{Gas and stellar kinematics}
\label{GSK_2}

\begin{figure}
\begin{centering}
\includegraphics[width=60mm]{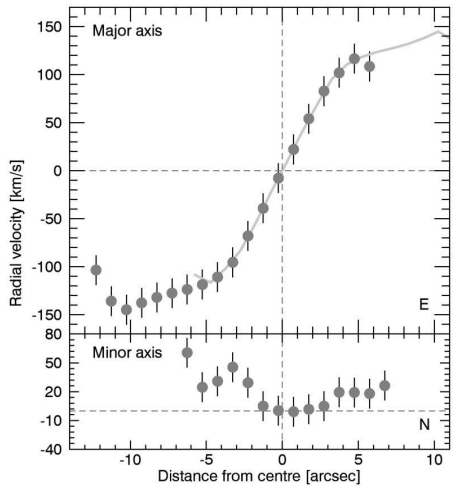}
\end{centering}
\caption{{\bf SOS\,90630} Gas radial velocity profiles along the main
  axes. The grey curve in the top panel is the flipped profile plotted
  to assess its symmetry.}
\label{SOS90630gasprof}
\end{figure}

The velocity fields of the gas in SOS\,90630 are shown in
Fig.~\ref{SOS90630kin}, while Fig.~\ref{SOS90630gasprof} shows the
profiles of the gas radial velocity along the major and minor axes in
the disk of the galaxy. The kinematic axis (traced by the green
colour) is distorted into a U-shape with a concavity directed to the
West and is not symmetric with respect to the major axis.

The kinematics of the external gas is continuous with that of the
disc. The velocity in the tail smoothly rises from
$\sim-$145\,km\,s$^{-1}$ in the disk to $\sim-$20\,km\,s$^{-1}$ at the
extreme west edge (41\,kpc from the centre). Within 5\,kpc, where the
gas is present on both sides of the disk, the radial velocity is
symmetric with respect to the centre (Fig.~\ref{SOS90630gasprof},
upper panel). Only in the last kpc in the western disk we notice an
inversion of the trend. The minor axis profile just reflects the
U-shaped kinematic axis.

The velocity dispersion of the gas is fairly uniform in the disk,
around $\sim$60-80\,km\,s$^{-1}$, higher ($\sim$80-120\,km\,s$^{-1}$)
in the first part of the tail (17 to 21\,kpc from the centre) and
lower (around 50\,km\,s$^{-1}$) in the rest of the tail. In the disk,
we notice that the region with higher values of $\sigma$ is also
U-shaped like the radial velocity field.

\begin{figure} \begin{centering}
\includegraphics[width=70mm]{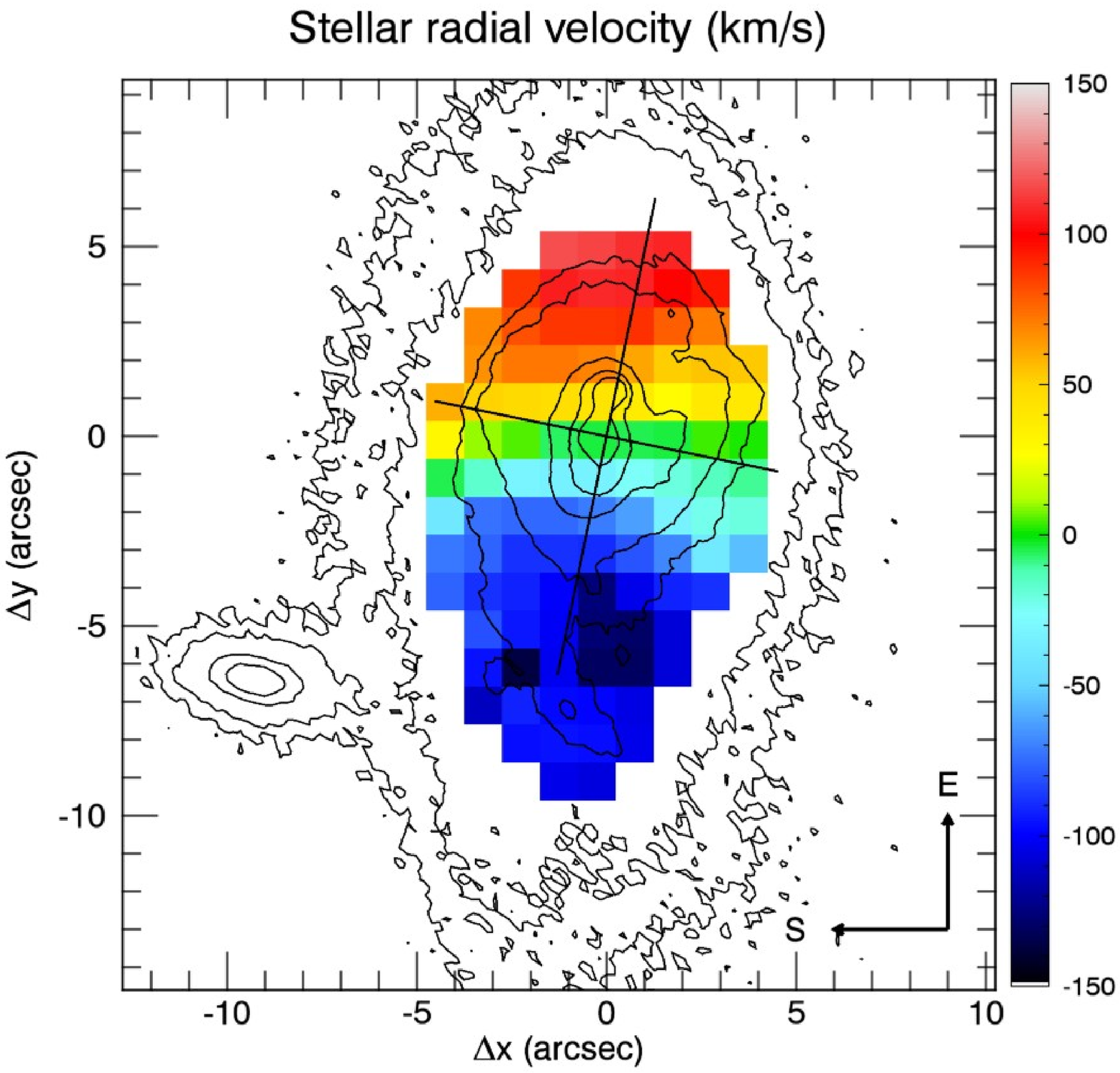}
\includegraphics[width=60mm]{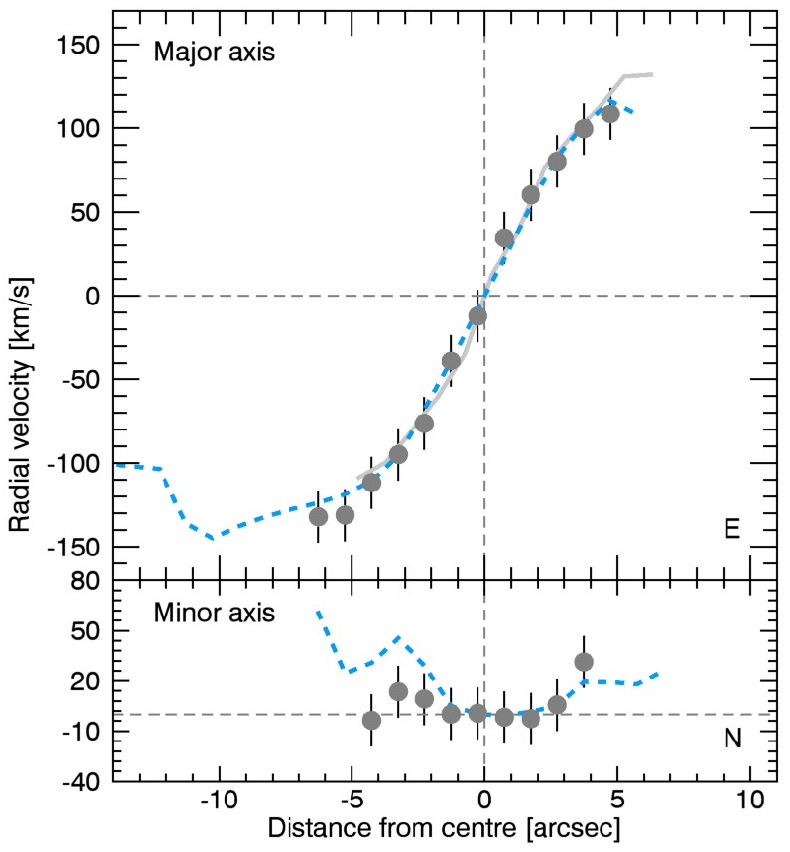} 
\end{centering} 
\caption{{\bf SOS\,90630} Stellar velocity field (top) with stellar radial
  velocity profiles (grey dots) compared with the gas one (cyan dashed
  line, bottom).}
\label{SOS90630sk} 
\end{figure}

The stellar radial velocity field and profiles, shown in
Fig.~\ref{SOS90630sk}, are almost fully consistent with those of the
gas along the major axis, but in the southern part of the minor axis
the gas recedes with a velocity higher (by $\sim$20\,km\,s$^{-1}$)
than the stars, indicating that the two components are decoupled
outside of the very central ($\lesssim 2$\,kpc) region of the galaxy.

\subsection{Dust extinction across the galaxy}
\label{SF_2}
\begin{figure}
\begin{centering}
\includegraphics[width=60mm]{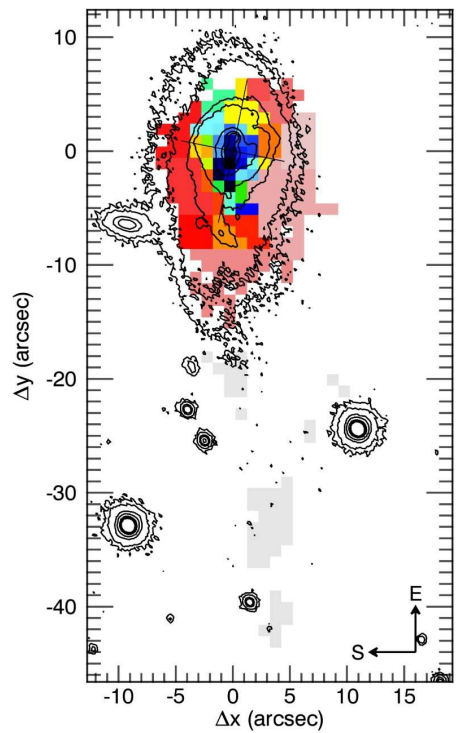}
\end{centering}
\caption{{\bf SOS\,90630} The 48 galaxy regions identified by the WVT
  are shown in different colours, which are consistently used in 
  Fig.~\ref{SOS90630KBPT}.}
\label{SOS90630bins}
\end{figure}

Similarly to the analysis of SOS\,61086, we performed a spatial
binning of the IFS data to reach the SNR needed for reliable
line-ratio measurements. As before, only the pixels with
SNR(H$\alpha$)$>8$ were considered in the analysis. The 48 regions so
obtained are shown in Fig.~\ref{SOS90630bins}. Similarly to
SOS\,61086, the continuum emission (from $u$ to $i$ bands), the
presence of compact H$\alpha$ emission regions and the diagnostic
diagrams suggest that the detected emission is mainly associated to
star formation. In Fig.~\ref{SOS90630SF} (upper panel) we show the
spatial distribution of the dust attenuation. This has the maximum (at
$\sim 2.4$\,mag) in the centre and is enhanced in the NW sector of the
disk ($\sim 1.5$\,mag), while decreases in the SE sector ($\sim
0.9-1.2$\,mag). The attenuation in the gas tail ($\sim 0.9$\,mag)
indicates a small dust content. The inset of Fig.~\ref{SOS90630SF}
(upper panel) is a zoom in the centre of the galaxy to show that the
western side is more attenuated (by $\sim 0.8$\,mag) than the eastern
side in support of the fact that the optical appearance of a double
nucleus is an artefact from dust absorption (see
Appendix~\ref{DN_app}).

\subsection{Ongoing star formation and recent star formation history}
\label{SOS90630SSH}

\begin{figure}
\begin{center}
\includegraphics[width=68mm]{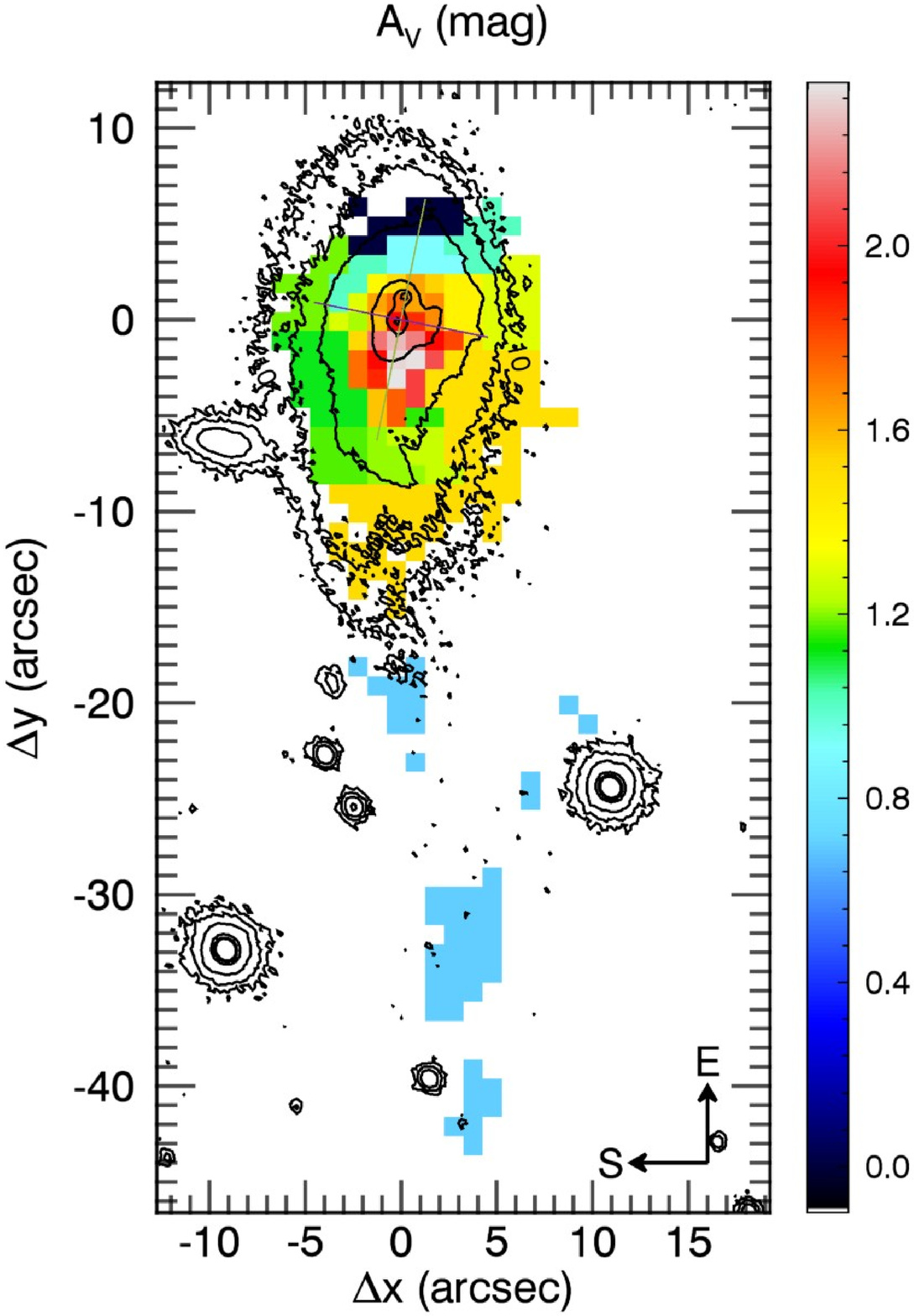}
\includegraphics[width=68mm]{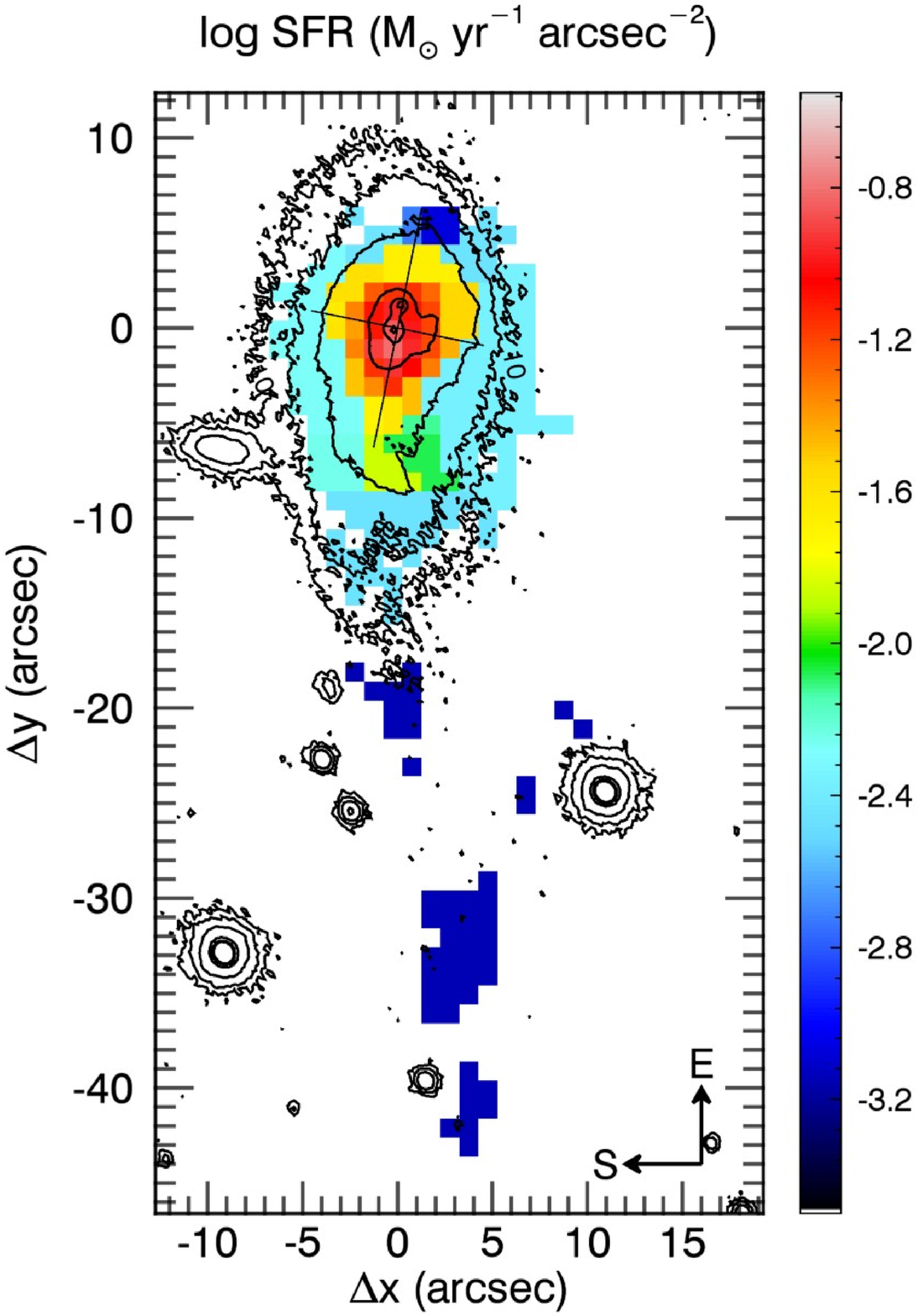}
\end{center}
\caption{{\bf SOS\,90630} Dust attenuation derived from the
  H$\alpha$/H$\beta$ line ratio (top) and attenuation-corrected SFR
  (bottom). The inset in the top panel is a zoom of the central region
  of the galaxy (see text).}
\label{SOS90630SF}
\end{figure}

The map of SFR in Fig.~\ref{SOS90630SF} (lower panel) shows that the
highest star formation takes place close to the galaxy centre
extending across the eastern clumps of H{\sc ii} regions and is then
truncated on the eastern boundary of the galaxy. The western spiral
arm in the H$\alpha$ imaging is also recognizable in this map. The SFR
drops abruptly in the external disk and is low in the tail. The
integrated H$\alpha$-derived SFR of SOS\,90630 (adopting the Kroupa
IMF) amounts to $3.49\pm 1.07$\,M$_\odot$yr$^{-1}$ where the error is
estimated as for SOS\,61086. The integrated H$\alpha$-derived SFR is
consistent within the errors with that derived from UV and IR data
(see Table~\ref{gals}).

The stellar continuum from the centre of SOS\,90630 appears relatively
flat from data gathered in the blue arm of WiFeS
(Fig.~\ref{fspect_SOS90630} upper panel), with 45\% of the integrated
emission at 4000{\AA} coming from stars formed less than 1\,Gyr ago
(thin blue curve), while the remainder comes from an old ($>$1\,Gyr)
stellar component. This older component appears primarily due to
intermediate age stars (1.5--3\,Gyr old) with little contribution from
stars formed at $z{>}1$. The young stellar component primarily
consists of a single burst of star formation extending over the last
200\,Myr, contributing 4\% of the stellar mass. This is still ongoing
as manifest by the strong H$\alpha$ emission with EW$({\rm
  H}\alpha){=}63${\AA} and other Balmer emission lines visible up to
H$\zeta$. This relatively mild recent increase in star formation is
also supported by the Ca{\sc ii} index of 0.81 and H$\delta/Fe${\sc
  i}$\lambda$4045 index of 0.76, both higher than those seen in
SOS\,61086, and consistent with a 0.3\,Gyr burst of star-formation
contributing 40\% of the integrated light at 4000{\AA}
\citep[c.f. Fig~3a of][]{LR96}. The stellar mass contribution of
recently formed stars ($<$200\,Myr old) increases slightly to 6\%
along the eastern edge, as does the strength of the H$\alpha$
emission.

\subsection{Excitation and Metallicity of the Gas}

\begin{figure}
\includegraphics[width=85mm]{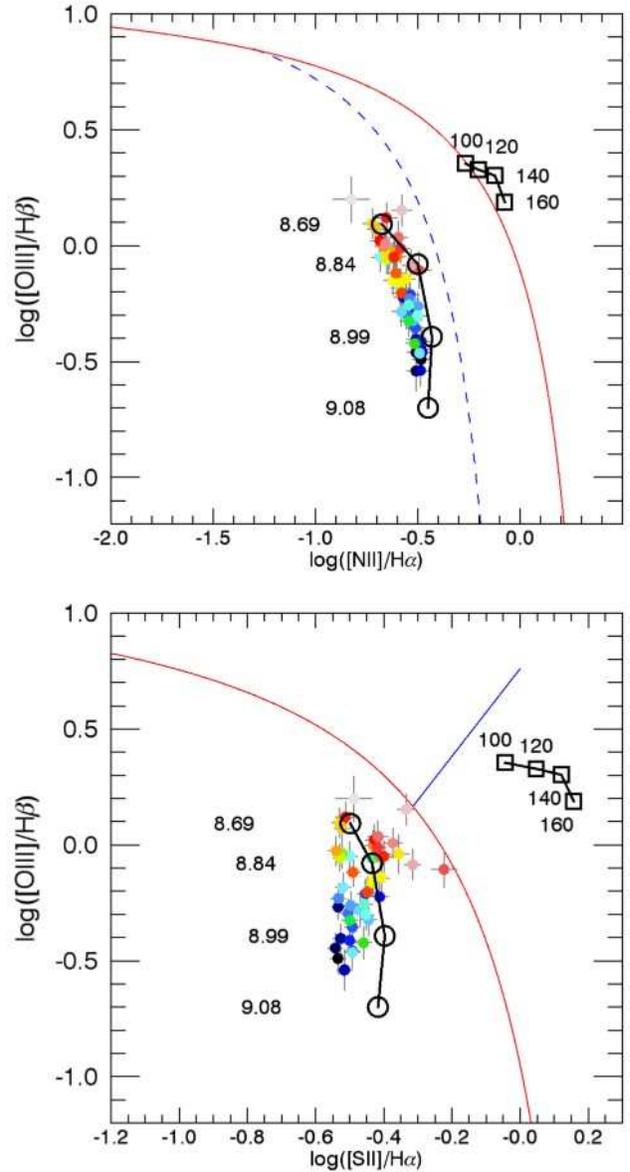}
\caption{{\bf SOS\,90630} Line flux diagnostic diagrams for the
  different regions of SOS\,90630. Colours code the galaxy regions
  shown in Fig.~\ref{SOS90630bins}. Other symbols like in
  Fig.~\ref{SOS61086KBPT}.}
\label{SOS90630KBPT}
\end{figure}

\begin{figure}
\begin{centering}
\includegraphics[width=70mm]{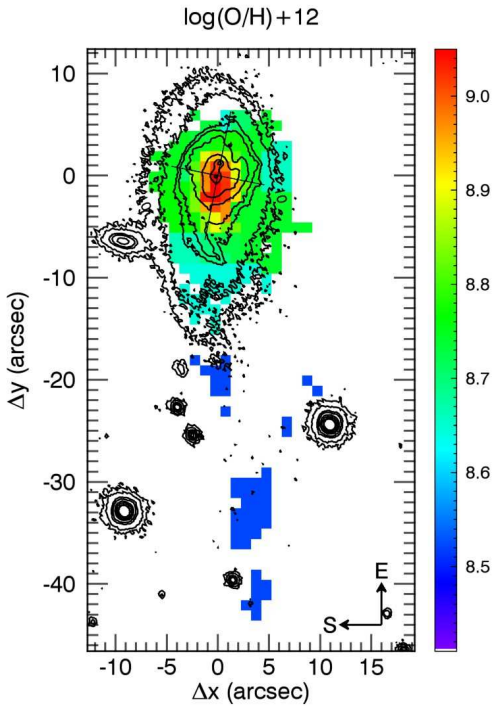}
\end{centering}
\caption{{\bf SOS\,90630} The O/H abundance ratios derived using the
  {\tt Pyqz} module.}
\label{SOS90630abu}
\end{figure}

The BPT diagrams for SOS\,90630 present no evidence that shocks play a
role in the excitation of the gas within the main body of the galaxy,
or within its extended tail. Indeed, the sequence of observed points
in Fig.~\ref{SOS90630KBPT} indicating that the excitation is due to
\HII\ regions with a fair range of abundance values. This abundance
range is caused by the fact that this galaxy has a clearly-defined
abundance gradient, as can be seen in Fig.~\ref{SOS90630abu}, where
the individual abundances are again derived using Python module {\tt
  Pyqz} described in \citet{Dopita13}. Also for this galaxy, the
ionisation parameter spans a narrow ($\sigma_{\log{q}} = 0.08$) range
of values around $\log{q}= 7.02$. The tail of the galaxy shows
appreciably lower oxygen abundance that the main body, identifying
this gas as having come from the outer regions of the galaxy.

\section{Discussion}
\label{epg}

SOS\,61086 and SOS\,90630 are characterized by extraplanar ionised gas
extending out to projected distances of 30 and 41\,kpc,
respectively. Such tails have been observed in other cluster
  galaxies \citep[e.g.][]{VHv05,SDV07,YKY07,YYK10,YYK12,BCF16}, but
their origin is not easy to determine since different possible causes,
acting at different epochs, may contribute to the observed
features. Such large-scale outflows of gas from a galactic disk may be
due to TI or RPS. We aim to investigate which of these mechanisms
dominates the evolution of the two galaxies although the solution may
not be unique.
 
Both galaxies have other supercluster members in their proximity
which could potentially interact with them. The best candidates for TI
with SOS\,61086 and SOS\,90630 are SOS\,61087 and SOS\,90090,
respectively.

\subsection{SOS\,61086}
The `companion' of SOS\,61086 is an early-type galaxy located at a
distance of $\sim$17\,kpc in projection and $\sim$300\,km\,s$^{-1}$ in
LOS velocity from SOS\,61086. The difference in redshift is lower than
the velocity dispersion of the host cluster SC\,1329-313. Since the
two galaxies have comparable masses, a possible interaction could
strongly affect their structures and SFs and may induce gas/star
outflow. However, we do not observe any clear disturbance in the
stellar disk of SOS\,61086 and SOS\,61087 is a `red and dead' galaxy
without any signs of perturbation.

The gas from the outer disk of SOS\,61086 has almost all been removed
from the outer edge of the disk beyond 7-8\,kpc from the centre, and
all that is left is the denser ISM close to the nucleus and that
associated with the inner disk. The truncated gas disk and the
fan-shape of the extraplanar gas points to ongoing RPS. The gas tail
expands to the North in the direction of the possible companion
galaxy, but it extends more than 10\,kpc further in projection beyond
SOS\,61087 in such a way that it cannot be a bridge between the two
galaxies. We already noticed that the rotational component of the gas
disk is preserved in the extraplanar material (at least up to 12\,kpc
from the disk) as seen in other cases of galaxies subjected to
ram-pressure \citep[e.g.][]{ACCESSV,FFH14}. No tidal stripping
scenario would be able to preserve such a rotational signature in this
way, while RPS acting almost face-on could provide a natural
explanation.

The gas velocity profile is asymmetric as foreseen in galaxies
affected by RPS and TI \citep{KKU08,KKS06}, but a comparison of the
gas and stellar velocity profiles shows that the two components are
clearly decoupled with the gas having lower velocities on average and
the stellar profile being fairly symmetric. This strongly favours the
RPS hypothesis.

The peak of luminosity in the centre is dominated by a young
population ($<$1\,Gyr) of stars indicative of a recent/ongoing
starburst. This feature can be associated to RPS as shown by
simulations \citep{KSS09,SHK12} and observations
\citep{ACCESSV,KGJ14,YGF13}. \citet{CK08} conducted a survey of 10
Virgo galaxies with SparsePak IFS. They found clear evidence of
RPS-induced star formation within the truncation radius and a passive
population beyond it. A burst of star formation can also be induced by
TI \citep[e.g.][]{KKS06}, but would take place close to the pericentre
passage \citep[e.g.][]{DBM08}, where the galaxies are heavily
perturbed, and this is not our case. The top panel of
Fig.~\ref{Ha61086} shows H$\alpha$ emission knots well out of the disk
and also this feature can be related to both RPS and TI. With respect
to the source of excitation, differentely to other galaxies affected
by ram-pressure, here shocks are not important: the line ratios are
characteristics of H{\sc ii} regions across the galaxy and in the
tail, except for one region in the NE tail. This is not the case in
other galaxies showing RPS such as NGC\,4330, NGC\,4402, NGC\,4501,
NGC\,4522 \citep{WKM14} and NGC\,4569 \citep{BCF16} in the Virgo
cluster, IC\,4040 \citep{YYK12} in the Coma cluster, ESO137-001 \citep
{FFB16} in the Norma cluster and SOS\,114372 \citep{ACCESSV} in the
cluster A\,3558. Nevertheless, our observations suggest that star
formation is occurring in the tail which is also inferred, for
instance, by \citet{YGF13} for NGC\,4388 in the Virgo cluster and
\citet{FFH14} for ESO137-001 in the Norma cluster.

In summary, the observational evidences support in general the RPS
scenario, although we cannot exclude a TI in a very early phase. We
investigate if RPS alone can explain the observed gas kinematics
running N-body/hydrodynamical simulations of RPS for SOS\,61086.

\subsection{SOS\,90630}
SOS\,90630 is distant 68\,kpc in projection and
$\sim$320\,km\,s$^{-1}$ in LOS velocity from SOS\,90090 and the two
galaxies belong to SC\,1327-312 which has a velocity dispersion of
535\,km\,s$^{-1}$. The tail of extraplanar gas is directed toward the
other galaxy (see Fig.~\ref{Ha90630}), which, incidentally, is a
factor 4-5 more massive.

The star formation across SOS\,90630 is highly asymmetric (upper panel
Fig.~\ref{Ha90630} and lower panel Fig.~\ref{SOS90630SF}) with a
`crown' of H{\sc i} regions tracing the eastern disk and other
star-forming regions in the western disk elongated toward the tail in
a spiral arm. A single burst of star formation over the last 200\,Myr
and still ongoing is responsible of the young stellar population
detected in the centre (Sect.~\ref{SOS90630SSH}). What induced this
burst and the asymmetric SF across the disk?

The gas and stellar velocity profiles coincide in the inner disc, but
are decoupled in the outer disk along the minor axis. Along the major
axis we observed the gas truncation in the leading  eastern edge and the gas
tail leading West following the galaxy rotation with an inversion only
in the more distant section. In case of RPS this would be acting
almost edge-on to reproduce the shape of the outflowing gas. The gas
emission is consistent with photoionisation throughout the galaxy and
in the tail.

The crown of H$\alpha$ emission evokes the decoupled dust clouds
observed in two Virgo spirals by \citet{AK14}. They show the bulk of
the dust being pushed back into the galaxy by ram-pressure, leaving
behind a population of isolated giant molecular clouds located up to
1.5\,kpc beyond the edge of the main dust lane. These molecular clouds
are the only parts of the ISM currently able to resist the ram
pressure which is acting to decouple them from the rest of the
lower-density ISM material. Similar features are also observed by
\citet{YGF13} in NGC\,4388. This is confirmed in the present case by
the low dust extinction measured in the leading edge and increasing
toward the galaxy centre and in the western galaxy disk (see
Fig.~\ref{SOS90630SF}). We notice however that similar features are
also found in hydrodynamical simulations of mergers \citep{TCB10}
where the gas response to the interaction is dominated by
fragmentation in dense clouds along the spiral arms.

The observational evidence alone does not allow us to establish which
mechanism prevails over the other. While the truncation of the gas
disk supports the RPS, other indicators as the $r$-band imaging tend
to TI between SOS\,90630 and the close massive SOS\,90090 although its
disk does not present obvious signs of perturbations. This reminds the
case of NGC\,4438 in the Virgo cluster with a stellar tidal tail and
extraplanar gas \citep{VBC05}, its companion NGC\,4435 presents an
almost regular disk in the optical image. We chose to run solely
hydrodynamical simulations of RPS even in this case for two main
reasons. Firstly, our aim is to investigate if RPS can play any role
and at what extent. Secondly, to set the hydrodynamical simulations of
TI would imply to make assumptions on the impact parameters, the
spatial trajectories and the relative velocities of the galaxies,
which are almost impossible to constrain, and therefore render the
results of such simulations very uncertain.

\subsection{Ram-pressure stripping} 

The effects of RPS depend on the properties of both the galaxy and the
ICM as well as the relative velocity. For the galaxies we adopt the
parameters listed in Table~\ref{gals} and Sect.~\ref{targets}. The ram
pressure is given by $P_{ram}{=}\rho_{ICM}V^2$, where $\rho_{ICM}$ is
the density of the ICM and $V$ is the velocity of the galaxy relative
to the ICM. These parameters are estimated as follows.

\subsubsection{ICM densities}
\label{ICMpro}

\begin{figure}
\begin{center}
\includegraphics[width=70mm]{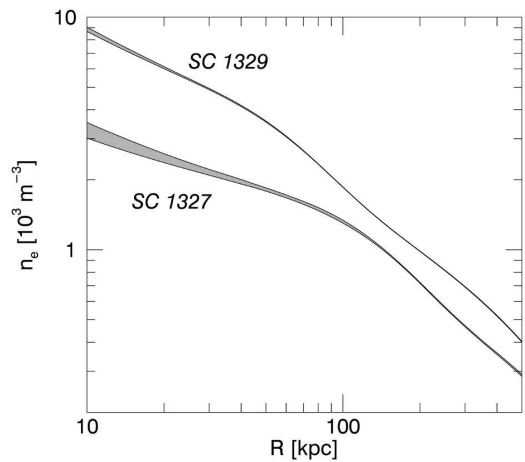}
\end{center}
\caption{Electronic density profiles centred on the X-ray surface
  brightness peaks of the galaxy clusters, SC\,1329-313 and
  SC\,1329-312 (see text). }
\label{edp}
\end{figure}

Successive XMM-Newton observations have now continuously covered the
SSCC filament that connects A\,3562, SC\,1329-313, SC\,1329-312 and
A\,3558. Out of these observations, photon-event light curves have
been analysed in order to filter any flare contamination,
i.e. observation periods with an anomalous count rate. This `cleaning'
procedure yields 11 filtered event-lists that we reprocessed using the
XMM-Newton Science Analysis System (SAS) version 14.0.0 in order to
select the relevant calibration files. Following a procedure described
in \citet{BM08}, the filtered event-lists have been re-binned in sky
coordinate and energy, and associated to both a 3D effective exposure
and a background noise model. As detailed in \citet{BMM13}, the
background noise model includes astrophysical and instrumental
components that have been jointly fitted within a sky area that
excludes the brightest region of each cluster. A wavelet denoised
image of the SSCC filament derived from this data set is shown in
Fig.~\ref{galsX}. Assuming that the ICM is spherically symmetric in
the innermost regions of SC\,1329-313 and SC\,1329-312, we deprojected
a couple of gas density profiles centred on each cluster peak. To do
so, analytic profiles of the gas density and temperature \citep{VKF06}
were projected along the line of sight following the weighting scheme
proposed by \citet{MRM04}, and fitted to the radially averaged surface
brightness and spectroscopic temperature. In this procedure, the ICM
emissivity is modelled following the Astrophysical Plasma Emission
Code \citep{SBL01}, shifted to $z=0.049$ and absorbed assuming a
Galactic hydrogen density column of N$_h = 3.96 \times
10^{20}$\,cm$^{-2}$. The derived profiles are shown in Fig.~\ref{edp}.

SOS\,61086 and SOS\,90630 are located at 282\,kpc and 226\,kpc,
respectively, in projection from the X-ray cluster
centres. Interpolating the density profiles at these radii, we obtain
electron densities $n_e$ of $\sim 5\times 10^{-4}$\,cm$^{-3}$ and
$\sim 9\times 10^{-4}$\,cm$^{-3}$, respectively. Assuming a
metallicity of Z=0.3 $Z_{\odot}$ and the element abundances of
\citet{GS98}, the particle mean weight per hydrogen atom is $\muh =
1.34732$, and the number of electrons per hydrogen atom is $\nuh =
1.17366$. Given these values, the ICM mass density can be derived from
the electronic density:

$$\rho_{\rm ICM} = \frac{\muh m_p}{\nuh} n_e\ \ \ .$$

\noindent
Therefore, electronic densities measured for SOS\,61086 and
SOS\,90630, at the projected distances, translate in gas mass
densities of $\rho_{\rm ICM}({\rm SOS\,61086})\simeq 9.6\times
10^{-28}$\,g\,cm$^{-3}$ and $\rho_{\rm ICM}({\rm SOS\,90630})\simeq
1.7\times 10^{-27}$\,g\,cm$^{-3}$. However, these are upper limits
since if the galaxies are at larger 3D distances, their local
densities will be lower. We account for this possibility considering
also lower values for the ICM densities. At about 2$\times$ the
projected radius we measure $\rho_{\rm ICM}({\rm SOS\,61086})\simeq
5.5\times 10^{-28}$\,g\,cm$^{-3}$ and $\rho_{\rm ICM}({\rm
  SOS\,90630})\simeq 7.7\times 10^{-28}$\,g\,cm$^{-3}$.

\subsubsection{Orbital velocities of the galaxies}
\label{wind}

As lower limits for the orbital velocities, we assume the galaxies'
LOS velocities relative to the clusters' systemic velocities
$V_{los,SOS\,61086}\sim$ -612\,km\,s$^{-1}$ and
$V_{los,SOS\,90630}\sim$ -336\,km\,s$^{-1}$ (both galaxies have lower
velocities with respect to the systemic ones, indicating that they are
moving toward the observer). The galaxies have also a velocity
component in the plane of the sky, as suggested by the geometry of the
gas tails. For the upper limits to the spatial velocity we assume a
velocity equal to 3$\times$ the cluster velocity dispersion,
i.e. 1000\,km\,s$^{-1}$ and 1500\,km\,s$^{-1}$ for SOS\,61086 an
SOS\,90630, respectively.

The inclination of the galaxy disk and the distribution of the
extraplanar gas suggest that the RPS is acting almost face-on on
SOS\,61086 and almost edge-on on SOS\,90630. The inclination angle
$\beta$ between the galaxy rotation axis and the ICM wind direction
has been shown to influence the amount of the gas pushed out of the
disk in the first phase of the stripping, only in case of moderate ram
pressure \citep[e.g.][]{MBD03,RH05}. On the other hand, $\beta$ can be
important in drawing the shape of the stripped gas. We note that in
the observed shape another factor is crucial, i.e. the position of the
observer with respect to both the wind and the galaxy disk.

For SOS\,61086 there is an indication that the galaxy is not moving
radially into the cluster: the extraplanar gas extends N-S while the
cluster centre is located West of the galaxy position projected in the
sky. We explored the values in the ranges: 500-1000\,km\,s$^{-1}$ and
500-1500\,km\,s$^{-1}$ for the orbital velocity of SOS\,61086 and
SOS\,90630, respectively. For the inclination angle, we considered a
wide range $0^\circ \leq \beta \leq 80^\circ$. Our approach was to run
low-resolution simulations for a grid of parameters and then select
the most promising cases for the high-resolution simulations.

\subsubsection{Constraints on RPS}

To check if the expected ram pressure is able to strip the gas from
the disk of SOS\,61086 and SOS\,90630, we compare it to the
gravitational restoring force per unit area
$(d\phi/dz)\Sigma_{gas}=2\pi G\Sigma_{star}\Sigma_{gas}$, with
$\Sigma_{star}$ and $\Sigma_{gas}$ the star and gas surface density,
respectively \citep[e.g.][]{KGV04,DMK06}. The stellar mass surface
density in the disk is given by
$\Sigma_{star}(r)=\Sigma_{star}(0)\times exp(-r/r_d)$ where $r_d$ is
the disk scale radius and $\Sigma_{star}(0)=\mathcal{M}_d/(2\pi
r_d^2)$ with $\mathcal{M}_d$ the mass of the disk. The disk scale
radii are $r_d=2.86$\,kpc and $r_d=2.05$\,kpc for SOS\,61086 and
SOS\,90630 respectively\footnote{These quantities were derived by
  fitting the VISTA $K$-band images using GALFIT \citep{PHI10} and
  corrected for dust absorption following \citet{GW08}.}

The mass surface density at the stripping radius, $r_{strip}$, is
$\Sigma_{star}(r_{strip})=\Sigma_{star}(0)\times
exp(-r_{strip}/r_d)$. The ram pressure is equal to the restoring force
at radius $r_{strip}$ when $\rho_{ICM}V^2=2\pi G
\Sigma_{star}(0)\times exp(-r_{strip}/r_d)\Sigma_{gas}$. We assume a
typical $\Sigma_{gas}\sim 10$\,M$_\odot$pc$^{-2}$
\citep[e.g.][]{KGV04}. For SOS\,61086 we have $\mathcal{M}_d=2.8\times
10^9$M$_\odot$, $r_d=2.9$\,kpc, $r_{strip}=6.2$\,kpc so the ram
pressure can be effective in removing the gas from the disk if $V\geq
350$\,km\,s$^{-1}$ and $V\geq 470$\,km\,s$^{-1}$ with high and low
density of the ICM (see Sect.~\ref{ICMpro}). For SOS\,90630 we have
$\mathcal{M}_d=9.5\times 10^9$M$_\odot$, $r_d=2.0$\,kpc,
$r_{strip}=6.1$\,kpc implying $V \geq 450$\,km\,s$^{-1}$ and $V \geq
650$\,km\,s$^{-1}$ for high and low density of the ICM (see
Sect.~\ref{ICMpro}). In these estimates we assumed the ram pressure
acting face-on but simulations show that the inclination angle does
not play a major role unless the ram pressure is acting very close to
edge-on \citep{RB06}.

Combining the ICM density estimates and the velocity ranges for the two
galaxies, the ram pressure ranges from $\sim 1.4\times
10^{-12}$\,dyn\,cm$^{-2}$ to $\sim 9.6\times 10^{-12}$\,dyn\,cm$^{-2}$
for SOS\,61086 and from $\sim 1.9\times 10^{-12}$\,dyn\,cm$^{-2}$ to
$\sim 3.8\times 10^{-11}$\,dyn\,cm$^{-2}$ for SOS\,90630. Following
\citet{RH05} these are the regimes of weak to moderate ram pressure
for SOS\,61086 and moderate to high ram pressure for SOS\,90630. Their
2D hydrodynamical simulations demonstrated that ram-pressure effects
can be observed over a wide range of ICM conditions and even in low
density environments, such as cluster outskirts and poor clusters as
in this case, where the moderate ram pressure can bend the gaseous
disk of $L^\star$ galaxies. Taking into account that our galaxies are
2.7-2.1\,mag fainter than $L^\star$, we expect the ram pressure being
much more efficient up to determining the truncation of the gas disk,
as we observe \citep[e.g.][]{MBD03}.

\section{Does ram-pressure stripping explain the observations?}
\label{sim}

The main goal of the simulations is to understand to what extent ram
pressure might explain the main observed features of the gas: the
truncation of the disk, the shape and extension of the tail and the
kinematics. This would eventually constrain the ICM wind angle and
velocity and the time of the onset (`age') of RPS.

\begin{table}
  \centering
\caption{{\bf Set-up of the simulations.}}
{\small 
 \begin{tabular}{l|cc}
    \hline 
     & {\bf SOS\,61086} & {\bf SOS\,90630} \\
    \hline 
    $c$ & 9.619 & 9.165 \\
    $\alpha$ & 1.556 & 1.071 \\
    $\beta$ & 2.691 & 2.797 \\
    $\gamma$ & 0.7897 & 0.993 \\
    $r_s [\mathrm{kpc}]$ & 7.91 & 9.08 \\
    $\rho_s\left[\mathrm{M}_\odot/\mathrm{yr}\right]$ & $0.705\times10^7$ & $1.161\times10^7$\\
    \hline 
    $v_{200} \left[\mathrm{km}/\mathrm{s}\right]$ & 93 & 120 \\
    $\lambda$ & 0.033 & 0.033\\
    $m_d$ & 0.0285 & 0.0355 \\
    $m_b$ & 0.00392 & 0.00369 \\
    $f_{\mathrm{gas}}$ & 0.48 & 0.24 \\
    $z_0$ & 0.2 & 0.2 \\
    $r_d \left[\mathrm{kpc} \right]$ & 2.61 & 2.05 \\    
    \hline
  \end{tabular}}
  \label{tab:gal_prop}
  \vspace*{0.2cm}
{\small
\begin{tabular}{l} 
Initial properties of the two galaxy models. First, the param-\\ eters
of the modified NFW \citep{DBD14} halos: con-\\ centration $c$;
shape parameters $\alpha$, $\beta$, $\gamma$; scale radius $r_s$;
scale \\ density $\rho_s$. Afterwards the setting for disk and bulge,
follow-\\ ing Springel et al. (\citeyear{Springel2005a}). $v_{200}$ is
the velocity at the virial ra-\\ dius $r_{200}$, $\lambda$ the
dimensionless spin parameter. The disk and \\ bulge mass fractions
are $m_d$ and $m_b$, with $f_\mathrm{gas}$ being the ini-\\ tial gas
amount in the disk and $z_0$ giving the disk height as \\ a fraction
of the disk scale length $r_d$.
\end{tabular}}
\end{table}

\subsection{N-body/hydrodynamical simulations of RPS}

We performed simulations of model galaxies resembling SOS\,61086 and
SOS\,90630 and experiencing RPS. The simulations have been done with
the cosmological moving-mesh code \texttt{AREPO} \citep{Springel2010}.
Contrary to ordinary grid codes, the Euler equations of ideal
hydrodynamics are solved on a moving mesh which allows to construct a
Galilean-invariant scheme. Such an approach allows an accurate treatment
of gas mixing and fluid instabilities (Rayleigh-Taylor or
Kelvin-Helmholtz instabilities), both crucial in RPS simulations. In
the simulations, radiative cooling \citep{Katz1996} and a sub-grid,
multi-phase model for star formation and stellar feedback
\citep{Springel2003} are included as well.

To simulate a galaxy undergoing RPS, we use a wind-tunnel setup
similar to \citet{Hess2012}. First, model galaxies are generated,
resembling pre-interaction properties of SOS\,61086 and
SOS\,90630. Initial conditions for an exponential stellar and gaseous
disk, as well as a stellar bulge with a \citet{Hernquist1990} profile,
of those model galaxies are calculated according to
\citet{Springel2005a}, based on theoretical work by
\citet{Mo1998}. For the dark-matter halo we are using a modified NFW
profile according to \citet{DBD2014}, and also include a hot gas
halo. The properties of the model galaxies are shown in
Table~\ref{tab:gal_prop}. Those model galaxies are put into a cuboidal
simulation domain with extension $200\times100\times100\,\mathrm{kpc^3}$
with the model galaxy statically placed at
$(50,50,50)\,\mathrm{kpc}$. The dark-matter halo is represented by a
static gravitational potential, yielding an additional force on
stellar particles and gas cells in addition to their self-gravity.

RPS is simulated by imposing a wind on the model galaxies. To this
extent, gas cells are inserted in front of the galaxy with the
preferred density, velocity and a constant volume of
$4\,\mathrm{kpc}^3$ or $40\,\mathrm{kpc}^3$, depending on the
resolution of the simulation. When they pass by the galaxy, the
resolution is adjusted and gas cells are refined in order to avoid too
large volume discrepancies of neighbouring cells. Furthermore, the
mass resolution of gas cells which mainly contain ISM, is kept at a
constant value in order to produce stellar particles with a constant
mass. At the end of the simulation domain, all gas cells are removed.
To distinguish between gas cells containing either ICM or ISM, a
colouring technique as described in e.g. \citet{Vogelsberger2012} is
used. Gravitational softening lengths are set according to
\citet{Hayward2014}. We are using $2\times10^5$ particles/cells for
the stellar and gaseous disk respectively, as well as $5\times10^4$
particles for the stellar bulge. On average, the wind tunnel contains
$4.5\times10^5$ gas cells. For the low-resolution runs (LR), we use a
tenth of the particles/cells.

The models were run from the onset of the ram pressure to 1\,Gyr after
and stored in steps of 10 Myr. During this time the ram pressure is
kept constant and its onset is immediate.

\begin{figure}
\begin{center}
\includegraphics[width=70mm]{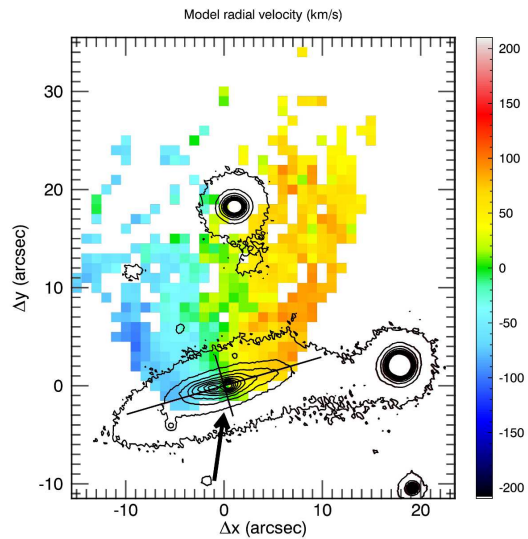}
\end{center}
\caption{{\bf SOS\,61086.} The best model ($\beta=30^{\circ}$,
  $V_{wind}$=750\,km\,s$^{-1}$ and $t=$ 250\,Myr) of the simulated gas
  velocity field with superimposed the $r$-band isophotes
  (dark-red). The projected wind direction is indicated by the white
  arrow. This figure should be compared with Fig~\ref{SOS61086kin}.}
\label{sim61086}
\end{figure}

\begin{figure}
\begin{center}
\includegraphics[width=70mm]{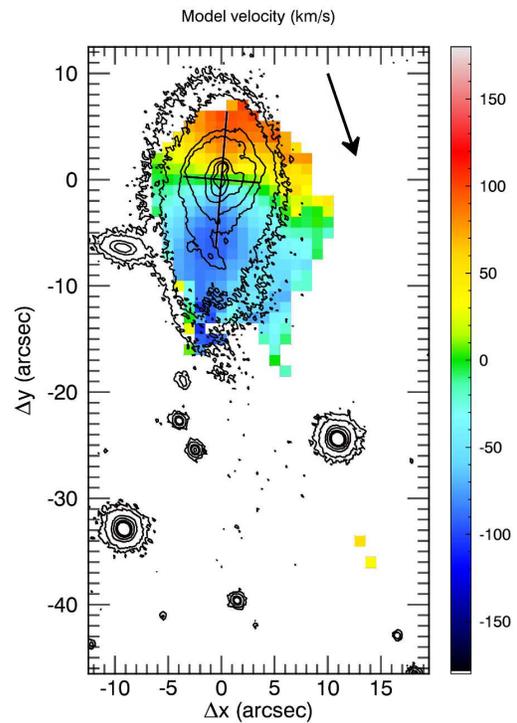}
\end{center}
\caption{{\bf SOS\,90630.} The best model ($\beta=80^{\circ}$,
  $V_{wind}$=500\,km\,s$^{-1}$ and $t=$ 120\,Myr) of the simulated gas
  velocity field with superimposed the $r$-band isophotes
  (dark-red). The projected wind direction is indicated by the white
  arrow. This figure should be compared with Fig.~\ref{SOS90630kin}.}
\label{sim90630}
\end{figure}

\subsection{Simulations analysis}

The simulations were run in the ranges of wind velocity and
inclination angle defined in Sect.~\ref{wind}. To compare the results
of simulations to the observed velocity fields, we proceeded as
outlined in \citet{ACCESSV} and summarized below. We first oriented
the models with respect to the observer in such a way that their LOS
velocity equals the observed velocity relative to the parent
cluster. Then, we integrated the velocities of the individual cells
along the LOS and binned the derived velocity fields in square pixels
of 1\,arcsec side, to simulate our IFS data. In doing this, we only
considered cells whose fraction of disk gas is greater than a certain
threshold to minimize the bias on the velocity field coming from ICM
particles. We adopted a threshold of 0.5, although the results remain
constant for a relatively wide range (0.2-0.9) of this parameter.  A
visual comparison of the data and the models then leads to the
selection of the parameters of the best RPS models.

In Fig.~\ref{sim61086} we show the simulation which better matches the
gas velocity field of SOS\,61086 and the projection on the plane of
the sky of the ICM wind direction. This model corresponds to
$\beta=30^{\circ}$, $V_{wind}$=750\,km\,s$^{-1}$, $t=$ 250\,Myr and
$\rho_{ICM}\simeq 5.5\times 10^{-28}$\,g\,cm$^{-3}$. We notice an
overall agreement between the model and the WiFeS observations
(Fig~\ref{SOS61086kin}, left panel). Both the velocity range and the
gas distribution are well reproduced by the model, where the gas
extends further in projection with respect to
SOS\,61087. Nevertheless, two particular features cannot be reproduced
by any models: the presence of the ionised gas in the central south
disk and the blue shifted gas clumps observed in the furthermost
extraplanar gas.

The neutral and ionised gas are assumed to be mixed as suggested by
simulations and observations \citep{TB10,AKC11}, but in principle the
simulations show the velocity field of the whole gas, while our data
refer to the ionised gas. This can be the origin of discrepancies
between models and observations. In particular, the ionised gas
in the south disk may be due to a galactic wind or the presence of
denser clouds more resilient to the the ram pressure. The blue-shifted
furthermost clumps of gas detected by WiFeS suggest a scenario where
the disk screens off the detached gas which, being still bound, falls
back into the disk. Such a complicated configuration is difficult to
be reproduced by the simulations.

In the case of SOS\,90630, we obtained the best match for
$\beta=80^{\circ}$, $V_{wind}$=750\,km\,s$^{-1}$ and $\rho_{ICM}\simeq
7.7\times 10^{-28}$\,g\,cm$^{-3}$, which is shown in
Fig.~\ref{sim90630} after 120\,Myr from the ram-pressure onset. For
this galaxy, we are not able with the simulations to reproduce both
the tail in its whole extent and the observed truncation of the gas
disk. When in the simulations the tail forms, the disk appears much
more truncated than observed and the gas velocity field is not
consistent with the observed one. Therefore RPS alone does not seem
sufficient to produce the tail, and the help of another force is
needed to pull the gas tail from the disk.

\citet[]{GBM01} and \citet[]{VHv05} studied two cases in which the ram
pressure was aided by TI in forming the gas tails of two galaxies in
A\,1367 and Virgo. The role of TI is, in their cases, to loosen the
restoring force of the galaxy thus making RPS more effective.
SOS\,90090, the massive galaxy close in projection to SOS\,90630,
could in principle induce such a tidal disturbance. To investigate
this possibility, we estimate the acceleration $a_{tid}$ produced by
SOS\,90090 on the ISM of SOS\,90630 and compare it with the
acceleration from the potential of SOS\,90630 itself, $a_{gal}$,
following \citet{VHv05}. We have
$$\frac{a_{tid}}{a_{gal}}=\frac{M_{90090}}{M_{90630}}
(\frac{r}{R}-1)^{-2} \ \ \ ,$$ where $R$ is the distance from the
centre of SOS\,90630 and $r$ is the distance between the galaxies
\citep{VHv05}. As a proxy of the mass ratio we use the stellar mass
ratio, but of course we don't know $r$. In the special case in which
projected and true distance coincide (both galaxies in the plane of
sky), the above equation would imply that the tidal force begins to
dominate over the restoring force at $R>$18\,kpc, which is the
projected distance where the gas tail begins. At three disc scale
radii ($R\sim$8.5\,kpc), the above ratio is 0.17, which shows that the
tidal force might be able to produce some perturbation also in the
disk, as is possibly observed in the structure of the external SE disc
(see Fig.\ref{SOS90630VST}).

This is a very crude approach, however, since i) it is most probable
that the distance between the galaxies is larger; ii) the above
formula is just a rough approximation; and iii) we definitely lack the
knowledge of the orbits of the two galaxies (e.g. they might have been
closer in the past). On the other hand, in the light of the cases
quoted above, TI does not need to be the only or dominant mechanism to
produce the tail, but only strong enough to `help' the ram pressure,
and this might well be the case unless the two galaxies are very
distant in space. We finally notice that the direction of the gas tail
is intermediate between the direction of the wind in the RPS
simulation and the direction of SOS\,90090, which could point to a
collaborative role of the two mechanisms.

We can therefore conclude that the phenomenology of the ionized gas is
explained in part by the ram-pressure acting close to edge-on and
pushing the gas in the NW direction. For what concerns the tail, we
can only argue that TI with SOS\,90090 could have acted as an aid to
the RPS.

The crown of star-forming regions suggests ram-pressure induced star
formation due to the gas compression at the leading edge where the gas
disc is truncated and the dust seems to be swept out. In fact, the age
of the RPS is consistent with the burst of star formation, started
200\,Myr ago and still ongoing. Star forming regions in the tail,
foreseen in the simulations, are inferred by the H$\alpha$ knots
detected up to 40\,kpc from the disc in the narrow-band
imaging. However, our data do not allow to determine if these clumps
origin from newly formed stars or stars stripped from the disc by the
possible combined action of RPS and TI.

\subsection{The role of cluster-cluster interaction} 

The dynamical analysis \citep{BZV94,BZZ98} and diffuse filamentary
X-ray emission \citep{BZM96,KB99,HTS99} showed that the five clusters
of the SSCC are interacting with recent and/or ongoing cluster-cluster
mergers.

\citet{FHB04} proposed a tidal interaction between SC\,1329-313 and
A\,3562 to explain the observed properties of the hot gas in A\,3562
-- the tailed shape of the X-ray emission associated with SC\,1329-313
as well as the sloshing of the A\,3562 core. In their scenario
SC\,1329-313 was initially flying North of A\,3562 in the western
direction and after passage at pericentre with A\,3562 deflected
South. This is also suggested by the X-ray structure of SC\,1329-313
(see Fig.~\ref{galsX}). Radio observations confirmed this scenario
\citep{GVB05}. The western extension of the radio halo in A\,3562 and
the direction of the emission of J1332-3146a, a radio galaxy 150\,kpc
NE of SOS\,61086 in projection, support the idea that merger-induced
turbulence is present in the region between the centre of A\,3562 and
SC\,1329-313. We further remark that the fan-shape tail radio emission
of J1332-3146a follows the same direction of the outflowing ionised
gas of SOS\,61086, which is also a radio source. This particular
similarity in shape and orientation between the two tails, although
having different origins, may witness for the first time that cluster
interactions trigger RPS events by perturbing the ICM.

SOS\,90630 is member of SC\,1327-312 which also may have interacted
with A\,3558 \citep[e.g.][]{RGM07}, but in this case a RPS event alone
cannot explain the observations and another mechanism such as TI
should be considered. It is interesting to notice that this galaxy can
be easily identified as a {\it jellyfish} galaxy, i.e. the `perfect'
candidate of ongoing RPS.

\section{Summary and Final Remarks}
\label{summary}

The present study investigates the ongoing transformations of two
low-mass galaxies in poor cluster environments with the aim of
understanding which are the mechanisms responsible. SOS\,61086 and
SOS\,90630 have been selected from deep subarcsec-resolution imaging
as promising candidates of ongoing gas stripping. Both have a possible
companion which may also suggest a tidal interaction. We notice that
this can often be the case in dense cluster environments, complicating
the identification of ongoing ram-pressure stripping events.

From the analysis of multi-band data and IFS observations complemented
by {\it ad hoc} hydrodynamical simulations we come to envision the
following scenarios.

\noindent {\bf SOS\,61086.} The ionised gas distribution (truncated
disc, 30\,kpc tail) and velocity field together with the properties of
the stellar component (unperturbed morphology and velocity field)
support the RPS scenario. Star forming regions are found in the
central disk, but also in the outflowing gas, as foreseen in case of
RPS \citep{RBO14}. In fact, the gas turns out to be manly photoionised
throughout the galaxy and its tail. The dust is pushed by ram
pressure, with the leading edge showing much lower attenuation.

Although some of these features can be suggestive also of TI, the
hydrodynamical simulations of RPS well reproduce the overall gas
velocity field with $\beta=80^{\circ}$, $V_{wind}$=750\,km\,s$^{-1}$
and $t=$250\,Myr. It is encouraging that the time of the onset of ram
pressure agrees with the age of the young stellar populations ($\sim
200$\,Myr). As in SOS\,114372 \citep{ACCESSV} we are very likely
observing ram-pressure induced star formation. All this supports a
scenario where ongoing RPS is the dominant mechanism at work.

\noindent
{\bf SOS\,90630.} The most distinctive features of this galaxy are the
very long (41\,kpc) tail of ionised gas and the truncated gaseous
disk.  The attenuation is lower on the leading edge of the disk and
star formation could be present in the tail. In this case the ram
pressure explains the truncation of the gas disk, is consistent with
the distribution of dust and star formation, while RPS and TI
with SOS\,90090 are responsible for the long gas tail.

We notice that both galaxies are members of interacting
clusters, suggesting that these systems can trigger RPS events by
perturbing the ICM. Finally, this work provides another clear proof
that RPS can take place in very different environments from cluster
cores to cluster outskirts, from rich clusters to poor ones, affecting
in various ways galaxies of different masses.

\subsection{Final remarks}

All simulations of RPS indicate that it should be very efficient in
quenching star formation of cluster galaxies, but its effects depend
on ICM and ISM properties, which may easily change locally, as well as
on the galaxy orbits. All together these parameters contribute to
determine case-by-case what fraction of gas is stripped from the
galaxy and over what time-scales. In fact, there are events where the
ICM-ISM interaction is not able to strip all the gaseous reservoir of
a galaxy and then definitely quench its star formation. Which is the
impact of RPS on the evolution of cluster/group galaxies?

At present only a few tens of ongoing RPS events have been ascertained
and studied in detail, most of these galaxies belong to nearby
clusters with only a few cases found in clusters at intermediate
redshift
\citep[e.g.][]{OLK05,CMR07,CK08,VSC10,YYK10,AK14,WKM14,FFB16,BCF16}. Thus,
the identification of new cases of galaxies experiencing RPS is
definitely important. One of the reasons for the small statistics are
the intrinsic and observational biases such as the short time-scales
and the observer's point of view with respect to the galaxy
trajectory. The latter particularly disadvantages the selection of RPS
candidates based only on photometric data which allow us to recognize
distorted morphology and to glimpse one-side extraplanar emission, but
cannot identify other features such as the truncation of the gas disk
and robust detection of extraplanar gas. In addition, this approach is
not able at all to disentangle between the effects of TI and RPS,
which might be similar.

Nevertheless, recently \citet{PFO15} and \citet{MER15} have undertaken
a systematic search in two samples of nearby and intermediate-redshift
clusters, respectively, using almost the same morphological
criteria. They inferred that the morphologically selected galaxies are
RPS candidates. Among the 13 galaxies of the Shapley supercluster
observed with WiFeS because of hints of extraplanar emission from
their images, 9 ($\sim 70$ per cent) actually presented extraplanar
ionised gas and only 4 ($\sim 30$ per cent) turned out to be
significantly affected by ram pressure. Thus, a certain caution is
needed before drawing conclusions on the basis of samples of
candidates. \citet{ESE14} introduced three criteria to identify RPS
candidates: disturbed morphology indicative of a unilateral external
force' ii) brightness knots and colour gradients suggesting bursts of
star formation; iii) evidence of tails. If we adopt the criteria of
\citet{ESE14} to identify RPS candidates in our spectroscopic sample
of supercluster galaxies (80 per cent complete at m$^\star$+3, $i =
17.6$), we find that 0.5 per cent follows all the Ebeling et al.'s
criteria in full agreement with the 0.6 per cent found by
\citet{ESE14}\footnote{Once accounted for the non cluster members,
  i.e. 57 per cent of the selected galaxies by \citet{ESE14}.}. If we
adopt at least two of the three Ebeling et al.'s criteria and apply
our success percentage of the 30 per cent, we expect to identify about
25 cases of ongoing RPS in the whole ShaSS survey (11 clusters
covering 23\,sqdeg) which would be a significant increase in the
number of known ongoing RPS events. Although such a large number is
supported by the hypothesis that RPS is triggered by cluster mergers
\citep{OCN12}, only a campaign of IFS observations (or HI for nearby
clusters) can confirm this indicative conjecture and this is the
project we are carrying out with WiFeS.

\section*{Acknowledgments}
We thank the referee for the comments and corrections which improved
the presentation of our work.  The authors thank Volker Springel for
providing the simulation code AREPO and Tiziana Venturi for her help
with the interpretation of the radio data. The optical imaging is
collected at the VLT Survey Telescope using the Italian INAF
Guaranteed Time of Observations and reduced by A. Grado and
L. Limatola. CPH was funded by CONICYT Anillo project ACT-1122. PM and
GB acknowledge financial support from PRIN-INAF2014: {\it Galaxy
  Evolution from Cluster Cores to Filaments} (PI B.M. Poggianti). MAD
acknowledges the support of the Australian Research Council (ARC)
through Discovery Project DP130103925, and he would also like to thank
the Deanship of Scientific Research (DSR), King Abdulaziz University
for additional financial support as Distinguished Visiting Professor
under the KAU Hi-Ci program.  DS acknowledges the support from the
Austrian Federal Ministry of Science, Research and Economy as part of
the UniInfrastukturprogramm of the Focal Point Scientific Computing at
the University of Innsbruck.  \bibliographystyle{mn2e}
\bibliography{biblio}{}

\appendix
\section[]{The origin of the double peak of luminosity in SOS\,90630}
\label{DN_app}

\begin{figure}
\includegraphics[width=80mm]{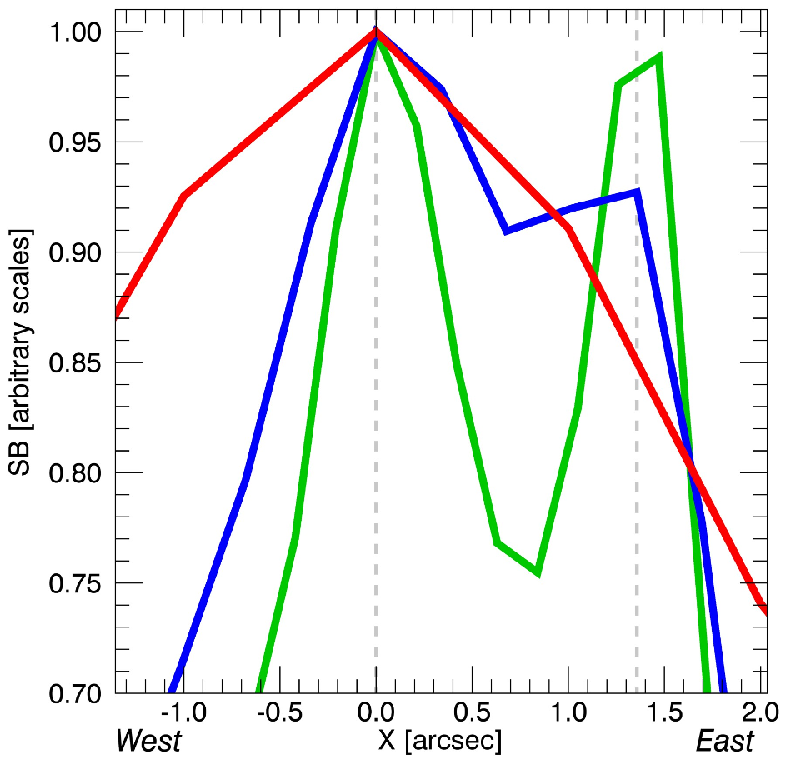}
\caption{Surface brightness (SB) profiles along the major axis of
  SOS\,90630 in the region of the two optical `nuclei'. The profiles
  in $r$ and $K$ bands are shown in green and red respectively. The
  blue curve is the profile of the $r$-band image after downgrading to
  the resolution and sampling of the $K$ band. The vertical dashed
  lines mark the positions of the two optical `nuclei'. The profiles
  are scaled in surface brightness to match in correspondence of the
  $K$-band (true) nucleus.}
\label{SOS90630nuclei}
\end{figure}

The optical ($gri$) images of SOS\,90630 in Fig~\ref{SOS90630VST}
present two peaks of luminosity in the centre suggesting the presence
of two nuclei which are not distinguishable in the $K$-band image. To
understand whether this might originate from dust absorption, we
downgraded the $r$-band image (FWHM=0.64\,arcsec,
0.21\,arcsec\,pxl$^{-1}$) to mimic the resolution and sampling in the
$K$ band (FWHM=1.1\,arcsec, 0.339\,arcsec\,pxl$^{-1}$).

Figure~\ref{SOS90630nuclei} shows the surface brightness profiles
along a line joining the two maxima in the $r$ band (incidentally,
this line coincides with the apparent major axis in the $K$-band image
of the galaxy). The green curve is the original profile in $r$ band,
the blue curve is the profile in the same band after downgrading, and
the red curve is the $K$-band profile. All profiles are arbitrarily
normalized to the maximum in $K$ band. It is clear that even with the
downgraded resolution the separation of the two `nuclei' persists in
the $r$-band profile, while in $K$ band there is instead only one
clear maximum with a fairly symmetric distribution around it. The
minimum of the $r$-band SB is located about 0.75\,arcsec East from the
nucleus. To explain the different profiles between optical and NIR, we
should also assume that the western side (left in the figure) is more
dust-attenuated than the eastern one. In Sect.~\ref{SF_2} we show that
this is the case. We therefore conclude that the appearance of a
`double nucleus' is caused by dust absorption.

\end{document}